\documentclass{article}
\usepackage{url}
\usepackage{amsmath}
\usepackage{amsfonts}
\usepackage{amssymb}
\usepackage{graphicx}
\usepackage{arxiv_new}
\usepackage{nicefrac}
\usepackage{microtype}
\usepackage{hyperref}
\usepackage{mathtools}
\usepackage{subcaption}
\usepackage{booktabs}
\usepackage{tabularx}
\usepackage{xcolor}

\makeatletter
\renewcommand{\@cite}[2]{(#1\if@tempswa , #2\fi)}
\renewcommand{\@biblabel}[1]{}
\makeatother

\begin{document}

\title{Rule-table structure of life-like cellular automata:\\
activity, asymmetry, sensitivity,\\
and the propagation--storage factorisation}
\author{Yair Cohen\\
Independent researcher\\
\texttt{yair.chn@gmail.com}\\
\url{https://www.linkedin.com/in/yair-cohen-a5a0a11a1/}}
\date{\today}
\maketitle

\begin{abstract}
What structure on the rule space of a cellular-automaton family exposes
which rules can support which behaviours? For the 262{,}144 life-like
rules---the two-dimensional outer-totalistic family containing Conway's
Game of Life---the standard coordinate, Langton's activity parameter,
is long known to be too coarse. I organise the family by three
symmetry-derived rule-table coordinates: an unweighted activity, a
birth--survival asymmetry (odd under the swap of birth and survival),
and the Vichniac--Binder Boolean-derivative sensitivity in closed form,
whose centre-bit component vanishes exactly on the centre-blind
($B=S$) submanifold. The cardinality coordinates ($\lambda'$, $\delta'$)
are necessary but never sufficient to certify behaviour; the centre-bit
sensitivity labels the $B=S$ submanifold on which the paper's one exact
exclusion holds. Combined with an exhaustive enumeration of every rule's
compact gliders, still lifes and oscillators (three-by-three seed,
period at most sixteen), they expose a hierarchical factorisation: the
birth condition is a narrow gate deciding whether anything can
propagate, the survival condition tunes storage, and every
compact-glider-supporting rule is birth-light and survival-rich.
One exclusion is exact: no $B=S$ rule supports a compact glider at that
footprint, proven by exhaustive enumeration and closing only for
compact-footprint patterns---36 $B=S$ rules ship at four-by-four seed
and 40 at five-by-five. Two dynamical results frame the whole:
glider-capable rules that random sampling never expresses show that
capability is certifiable from the table while its expression is only
estimable from dynamics; and every uncatalogued rule producing a richly
Life-like ecology turns out to be a Conway or HighLife variant on the
neighbour counts its trajectories actually visit.
\end{abstract}

\section{Introduction}

A cellular automaton (CA) updates a lattice of cells in parallel by a rule that
depends only on each cell and a small neighbourhood, yet even tiny rule families
produce propagating structures, long transients and universal
computation~\cite{vonneumann1966,wolfram1983,berlekamp1982}. This paper concerns
a basic organisational question for one such family: given the $262{,}144$
two-dimensional outer-totalistic ``life-like'' rules, what coordinates on the
rule table expose the structure of behavioural capability---which regions support
propagation, which support storage, and which exclude a behaviour outright? A
life-like rule
is fixed by two subsets $B,S\subseteq\{0,\dots,8\}$: a dead cell is born if its
live-neighbour count lies in $B$, and a live cell survives if its count lies in
$S$. The family contains Conway's Game of Life and its named variants---HighLife~\cite{bell1994},
Day~\&~Night, Seeds, $2{\times}2$, Replicator,
Maze~\cite{gardner1970,adamatzky2010,bays2010,eppstein2010}.

The standard coordinate is Langton's $\lambda$, the fraction of the rule table
that yields a live successor~\cite{langton1990}. Sweeping $\lambda$ moves through
frozen, periodic, complex and chaotic regimes, and the intermediate ``edge of
chaos'' was conjectured to host maximal computational capability~\cite{packard1988,langton1990}. But $\lambda$
is a single number summarising an $18$-bit table: two rules that differ in
\emph{which} neighbour counts fire, not just \emph{how many}, collapse onto the
same point, and evolved-CA studies found no sharp critical
$\lambda$~\cite{mitchell1993}. Prior work has organised the life-like family
along other single axes---Wuensche's $Z$-parameter sensitivity
ordering~\cite{wuensche1999}, P.-M.~Binder's $(\lambda,\mu)$
plane~\cite{binder1993}, and more recent classifications of the family's rule
space by dynamical-descriptor statistics~\cite{penasayama2021}---and
evolutionary searches have located individual complex rules
directly~\cite{sapin2003}. The present paper is complementary: rather than a single ordering axis
or an evolutionary search, it uses a small basis of symmetry-organised
coordinates paired with an exhaustive combinatorial enumeration of what each
rule can host at a fixed compact footprint.

\paragraph{Necessary, not sufficient.}
The collapse of rules that differ in \emph{which} neighbour counts fire onto
the same coordinate value is not a defect to be engineered away but a
structural fact worth stating plainly, because it sets the ceiling on what any
cardinality-based coordinate---one built only from how many elements are in
each set---can do.
Both $\lambda'$ and the asymmetry $\delta'$ I introduce below are functions of
the cardinalities $(|B|,|S|)$ alone; there are only $10\times10$
such pairs, so the entire $(\lambda',\delta')$ plane has at most $100$ occupied
points, each shared by thousands of behaviourally distinct rules. A coordinate
this coarse can never \emph{certify} that a rule does something---any value it
takes is shared by a crowd. What it can do is the opposite: place a rule in a
region, and---in the limiting case---\emph{exclude} a behaviour entirely. My
results are organised around this asymmetry. The asymmetry coordinate $\delta'$
locates where capability concentrates (a necessary condition); the centre-bit
sensitivity $\mu_{\text{centre}}$ supplies the one exact exclusion, ruling out
compact gliders on the centre-blind submanifold outright.

\paragraph{The coordinates.}
I add two complementary rule-table coordinates. The first is a
birth--survival asymmetry: Langton's $\lambda$ counts how much of the table
fires, and splitting that count into its birth and survival halves and taking
their difference measures whether a rule births more than it preserves or the
reverse. Built directly from Langton's parameter this asymmetry is $\delta$,
and it inherits Langton's multiplicity weighting---each neighbour count enters
weighted by how many neighbourhoods realise it. I also use unweighted siblings
$\lambda'$ and $\delta'$ that count the birth and survival conditions
themselves, without multiplicity. The structural results below are read on
this unweighted pair, with the Langton-consistent $\lambda,\delta$ kept as
reference; the specific empirical case for $\delta'$ over $\delta$ is
developed alongside those results (\S\ref{sec:coords}). Either way the asymmetry
is defined only for outer-totalistic families whose rule table separates by cell
state (dead-cell births $B$ vs.\ live-cell survivals $S$); it is a coordinate for
this family, not a re-parameterisation of $\lambda$ that generalises to CA whose
transitions do not factor this way.
The second coordinate is
the average Boolean-derivative \emph{sensitivity} $\mu$---the probability that
flipping one input bit of the neighbourhood flips the output---introduced for CA
by Vichniac~\cite{vichniac1990} and used as an a-priori ordering parameter by
Binder~\cite{binder1993}. For life-like rules $\mu$ admits a closed form whose
centre-bit part $\mu_{\text{centre}}$ vanishes exactly on the diagonal $B=S$.
The two new coordinates are independent, not interchangeable: under the
birth/survival swap $\tau\!:\!(B,S)\!\mapsto\!(S,B)$, $\delta'$ is odd while $\mu$
is even, so $\mu$ cannot tell a rule from its $B\leftrightarrow S$ mirror and
$\delta'$ can. With $\lambda'$ and the bit-flip complement $\sigma$ these three
coordinates populate three of the four parity sectors of $(\tau,\sigma)$; the
fourth is not used analytically here (\S\ref{sec:symm}).

\paragraph{Findings.}
Pairing this coordinate system with an exhaustive combinatorial enumeration of
every rule's compact-spaceship, still-life and oscillator support, I establish
three results. (i)~A hierarchical factorisation of capacity, read off the
birth--survival asymmetry $\delta'$: the birth set is a narrow gate that decides
whether anything can propagate, while the survival set tunes how much structure
is stored within that gate. The gate structure is robust across seed footprints
(the $3{\times}3$ propagation-enabling birth sets are a subset of the
$4{\times}4$ ones, both dominated by small birth sets); at the compact footprint
this shows up as a clean fingerprint---the rules supporting compact gliders are
systematically birth-light and survival-rich ($\delta'<0$)---which is the paper's
central structural finding, and which I show is specific to the compact footprint
and dilutes as the window widens. (ii)~Three coordinate projections of the rule
table---the Binder $(\lambda',\mu)$ plane and its $(\delta',\mu)$ and
$(\lambda',\delta')$ companions---that organise the family non-trivially, in which
the $512$ centre-blind rules occupy a sharply identifiable locus. (iii)~A
total exclusion of $B=S$ rules from compact ($3{\times}3$, period $\le16$)
gliders---certified by the exhaustive enumeration
(\S\ref{sec:disc-mechanism}). Two empirical regularities of the enumeration
organise the $B=S$ diagonal into a small candidate set that shipping rules
must inhabit: every $\phi>0$ rule satisfies $3\in B$ and $2\notin B$, so
$B=S$ shipping candidates collapse to $B=S=\{3\}\cup X$ with
$X\subseteq\{4,\ldots,8\}$---$32$ rules. This is an organising observation,
not itself a proof reduction, because $2\notin B$ is read from the same
exhaustive enumeration as the exclusion itself; a proof of the exclusion
from a mechanism independent of the enumeration remains open. The oscillator enumeration localises the exclusion to translation
rather than in-place recurrence, and it softens into a widening handicap
rather than vanishing at larger footprints. Two
dynamical results frame these. First, a population of rules that support a glider
combinatorially yet rarely or never express one under random sampling shows that
capability is a rule-table property, distinct from what random dynamics reveal
(auxiliary dynamical observables are correspondingly reframed as regime
descriptors, not capability detectors). Second, and conversely, a table-blind
screen of the uncatalogued build-and-hold rules for Game-of-Life--like dynamics
returns a richly Life-like ecology only for Conway and HighLife variants---rules
whose tables differ from an anchor's solely on neighbour counts their trajectories
never visit---so that table difference need not imply behavioural difference. The
exact bit-flip
duality $\sigma$ (\S\ref{sec:symm}) is used only to gauge-fix that auxiliary
dynamical sweep (\S\ref{sec:dyn}).

\section{Methods}

\subsection{Notation and computed quantities}
\label{sec:notation}

Table~\ref{tab:notation} collects every rule-table and dynamical quantity used in
the paper. The rule-table coordinates are defined in
\S\ref{sec:coords}--\S\ref{sec:symm}, the combinatorial indices in
\S\ref{sec:enum}, and the single dynamical sweep in \S\ref{sec:dyn}. All
rule-table quantities are functions of $(B,S)$ alone and are independent of any
simulation protocol.

\begin{table}[h]
\centering
\small
\renewcommand{\arraystretch}{1.25}
\begin{tabularx}{\linewidth}{@{}l >{\raggedright\arraybackslash}X >{\raggedright\arraybackslash}X@{}}
\toprule
Symbol & Definition & Measures \\
\midrule
$\lambda$ & $\bigl[\sum_{k\in B}\binom{8}{k}+\sum_{k\in S}\binom{8}{k}\bigr]/512$ & Langton's parameter: multiplicity-weighted fraction of the full table that fires \\
$\delta$ & $\bigl[\sum_{k\in B}\binom{8}{k}-\sum_{k\in S}\binom{8}{k}\bigr]/256$ & Langton-consistent birth--survival asymmetry \\
$\lambda'$ & $(|B|+|S|)/18$ & unweighted (condition-counting) activity; working coordinate \\
$\delta'$ & $(|B|-|S|)/9$ & unweighted birth--survival imbalance; the paper's factorisation axis \\
$\mu$ & avg.\ over inputs/bit-positions of $P[\text{flip flips output}]$ & sensitivity: how often flipping one input cell flips the rule's output \\
$\mu_{\text{centre}}$ & $P[n\in B\triangle S],\ n\sim\mathrm{Bin}(8,\tfrac12)$ & how much a cell's \emph{own} state matters to its fate; zero exactly when $B=S$ \\
$\phi$ & \# distinct $3{\times}3$ $D_4$-orbit spaceships, $T\le16$ & how many compact gliders the rule can support at all \\
$\psi$ & \# distinct $3{\times}3$ $D_4$-orbit still lifes ($T{=}1$) & how many small static patterns the rule can hold fixed \\
$\omega$ & \# distinct $3{\times}3$ $D_4$-orbit oscillators, $2\le T\le16$ & how many small patterns the rule can hold as in-place blinkers \\
$\mathcal{B}_{\text{prop}},\mathcal{B}_{\text{osc}},\mathcal{B}_{\text{sl}}$ & birth sets with at least one companion $S$ giving $\phi>0$ ($\omega>0$; $\psi>0$) & which birth rules ``open the gate'' for propagation / oscillation / storage \\
$\mathcal{G}^{*}_{\text{alt}}$ & peak off-axis residual spatiotemporal autocorrelation & whether random dynamics actually show a coherent moving structure \\
\bottomrule
\end{tabularx}
\caption{Notation and computed quantities. The first six rows are rule-table
coordinates (functions of $(B,S)$ alone); $\phi$, $\psi$, $\omega$ count
the small moving / static / blinking patterns a rule supports;
$\mathcal{G}^{*}_{\text{alt}}$ is the paper's single dynamical observable.
$B\triangle S=(B\setminus S)\cup(S\setminus B)$ is the
set of counts where birth and survival disagree; $\tau$ and $\sigma$ are the swap
and complement symmetries of \S\ref{sec:symm}; $D_4$ is the eight-fold symmetry
group of the square.}
\label{tab:notation}
\end{table}

\subsection{Rule-space coordinates: $\lambda$, $\delta$}
\label{sec:coords}

Each rule splits its look-up table into a birth component $B$ on dead cells and a
survival component $S$ on live cells. For a set $M$, $|M|$ denotes its
cardinality---the number of elements it contains---so $|B|$ is the number of
neighbour counts that trigger birth and $|S|$ the number that trigger
survival, each an integer in $\{0,1,\ldots,9\}$; for Conway $B=\{3\}$ and
$S=\{2,3\}$, so $|B|=1$ and $|S|=2$.
Langton's parameter is the fraction of the
full $2^9$-entry transition table that fires; on an outer-totalistic rule this
weights each neighbour count $k$ by its multiplicity $\binom{8}{k}$, and splits
by cell state into birth and survival halves,
\begin{equation}
  \lambda_B=\frac{1}{256}\sum_{k\in B}\binom{8}{k},\qquad
  \lambda_S=\frac{1}{256}\sum_{k\in S}\binom{8}{k},\qquad
  \lambda=\tfrac12(\lambda_B+\lambda_S),
\end{equation}
with antisymmetric companion
\begin{equation}
  \delta=\lambda_B-\lambda_S=\frac{1}{256}\Bigl[\sum_{k\in B}\binom{8}{k}-\sum_{k\in S}\binom{8}{k}\Bigr].
\end{equation}
This $\delta$ is the birth--survival asymmetry built directly from Langton's
$\lambda$.

Alongside these I use \emph{unweighted} siblings that count firing conditions
rather than multiplicity-weighted table entries---each of the nine neighbour
counts contributing equally:
\begin{equation}
  \lambda_B'=\frac{1}{9}\sum_{k\in B}1=\frac{|B|}{9},\qquad
  \lambda_S'=\frac{1}{9}\sum_{k\in S}1=\frac{|S|}{9},\qquad
  \lambda'=\tfrac12(\lambda_B'+\lambda_S')=\frac{|B|+|S|}{18},
\end{equation}
with antisymmetric companion
\begin{equation}
  \delta'=\lambda_B'-\lambda_S'=\frac{|B|-|S|}{9}.
\end{equation}
The two coordinates share their form and differ only in the summand: the
weighted $\lambda$ counts each firing count $k$ with the multiplicity
$\binom{8}{k}$ of neighbourhoods that realise it---the fraction of the full
$512$-entry table that fires---while $\lambda'$ gives every count weight one,
the fraction of the $18$ reduced outer-totalistic conditions that fire.
The weighted and unweighted forms agree only at the extremes and differ in
between---for Conway, $\lambda=140/512=0.273$ against $\lambda'=3/18=0.167$,
because the multiplicity weights concentrate on the middle counts that a bare
cardinality treats equally. The choice is not cosmetic: $\lambda$ and $\delta$
order rules differently from $\lambda'$ and $\delta'$, and the unweighted pair
discriminates compact-glider capability better (the birth-light propagation
family disagrees at low counts, which the binomial weighting under-emphasises;
$\delta'$ separates $\phi>0$ from $\phi=0$ at rank-biserial $-0.42$ against
$\delta$'s $-0.25$; the same ordering shows up under AUC ($0.71$ vs.\ $0.63$)
and holds direction at the $4{\times}4$ footprint too, though weakly ($0.56$
vs.\ $0.54$), consistent with the compact-fingerprint story fading with
footprint (\S\ref{sec:results-sigma})). No monotone rescaling relates the two,
since the difference is a reordering rather than a change of scale. I therefore
read the structural results on the unweighted $(\lambda',\delta')$, and retain
the weighted $(\lambda,\delta)$ as the Langton-consistent reference to which
the family connects. The $\phi>0$ population remains a single connected region
in the weighted $(\lambda,\delta)$ plane as well ($10$-vertex convex hull,
enclosing $\sim\!48\%$ of the $\phi=0$ bulk against $\sim\!78\%$ under the
unweighted coordinates); the weighted convex hull is thus \emph{tighter} as
a bounding envelope for $\phi>0$. On each 1D axis the weighted coordinate
takes $479$ distinct values across the $262{,}144$ rules against $19$ for
the unweighted (weighted counts every subset sum of $\{C(8,k)\}$; unweighted
counts $|B|-|S|$ only), so a scatter view of the weighted plane spreads the
$\phi>0$ population across many more discrete coordinate positions and
looks more diffuse pointwise, even while its convex hull is tighter overall.

\paragraph{$\delta'=0$ versus $B=S$.}
These loci are easily conflated but combinatorially distinct, and the distinction
is exactly the necessary/sufficient one. $\delta'=0$ is the equal-cardinality
submanifold $|B|=|S|$, containing $\sum_k\binom{9}{k}^2=\binom{18}{9}=48{,}620$
rules; the strictly diagonal $B=S$ (the centre-blind rules) is far smaller,
only $512$. This $B=S$ set is not new: it is the Life-like family's \emph{centre-blind}
subclass---rules whose update depends only on the neighbour count, not on
the centre cell's own state. It has sometimes been called ``totalistic'' in
the Life literature, but this differs from Wolfram's classical totalistic
convention (which sums over all $9$ cells including the centre); I use
``centre-blind'' throughout to avoid the ambiguity. What is new is the
characterisation of this subclass
as the exact zero set of the centre-cell sensitivity below, and the compact-glider
exclusion I prove on it. Equivalently, writing the unweighted set distance $|B\triangle S|$
(the number of counts in exactly one of $B,S$), $B=S$ is $|B\triangle S|=0$ while
$\delta'=0$ is only $|B|=|S|$, so $\{B=S\}\subsetneq\{\delta'=0\}$. The coordinate
that isolates $B=S$ exactly is the centre-bit sensitivity introduced next, of
which $|B\triangle S|$ is the unweighted shadow: as follows from the closed
form in \S\ref{sec:mu} (point~1), $B=S$ is precisely the $\mu_{\text{centre}}=0$
locus; \S\ref{sec:results-mu} confirms this empirically over the full family.

\subsection{Sensitivity coordinate $\mu$ (Boolean derivative)}
\label{sec:mu}

The $(\lambda',\delta')$ coordinates summarise activity but say nothing about how
strongly a rule responds to single-bit perturbations of its input. The standard
summary is the average Boolean derivative~\cite{vichniac1990,binder1993}: for a
rule with $N$ input bits and binary output $o$,
\begin{equation}
  \mu=\frac{1}{N\,2^{N}}\sum_{x\in\{0,1\}^N}\sum_{i=1}^{N}
  \bigl[o(x\oplus e_i)\neq o(x)\bigr],
\end{equation}
where $x$ runs over all $2^{N}$ binary configurations of the $N$-bit
neighbourhood (for life-like rules $N=9$: the centre cell and its eight Moore
neighbours), $e_i$ is the string with a single $1$ in position $i$, so
$x\oplus e_i$ is $x$ with exactly bit $i$ flipped ($\oplus$ is bitwise
exclusive-or), and the brackets $[\cdot]$ are the Iverson convention---$1$
when the enclosed condition holds and $0$ otherwise. The prefactor
$1/(N\,2^{N})$ averages the resulting flip-count over all
(configuration, bit-position) pairs, so $\mu$ is
the probability that a single-bit flip flips the output; it is the nearest
sensitivity-style ordering to Wuensche's $Z$ parameter~\cite{wuensche1999},
which similarly captures how much the update depends on individual input bits
but is defined on the full lookup table rather than as the closed-form
Boolean derivative for the outer-totalistic reduction I use below. For an
outer-totalistic life-like
rule on the Moore neighbourhood the $N=9$ inputs split as one centre bit and $8$
neighbour bits, and the table depends on inputs only through $(c,n)$ with $c$ the
centre value and $n$ the neighbour count. With $o(c,n)=\mathbf 1[c{=}0]\,\mathbf
1[n\in B]+\mathbf 1[c{=}1]\,\mathbf 1[n\in S]$, the centre-bit contribution
has the closed form
\begin{equation}
  \mu_{\text{centre}}=\frac{1}{2^{8}}\sum_{n=0}^{8}\binom{8}{n}\,
    \mathbf 1[n\in B\triangle S].
\end{equation}
Only the centre-bit part is used analytically below; the neighbour-bit
contribution has the analogous closed form
\begin{equation}
  \mu_{\text{neigh}}=\frac{1}{2^{8}}\sum_{k=0}^{7}\binom{7}{k}
    \bigl(\mathbf 1[k\in B\triangle(B{-}1)]+\mathbf 1[k\in S\triangle(S{-}1)]\bigr),
\end{equation}
where $B\triangle(B{-}1)$ picks out the counts at which set-membership
switches between $k$ and $k+1$ (the ``edges'' of $B$ as a subset of
$\{0,\ldots,8\}$). Then $\mu=(\mu_{\text{centre}}+8\,\mu_{\text{neigh}})/9$, a
weight-$1/9$/weight-$8/9$ average.
Giving the two sides
of the disagreement opposite signs turns $\mu_{\text{centre}}$ into a $\tau$-odd
quantity, $\mu_{\text{centre}}^{\text{odd}}=\tfrac1{256}[\sum_{n\in B\setminus S}\binom{8}{n}-\sum_{n\in S\setminus B}\binom{8}{n}]$;
the shared counts cancel, so this equals the weighted asymmetry $\delta$ of
\S\ref{sec:coords} identically. Signing a sensitivity thus reproduces the
Langton-consistent $\delta$, not the unweighted $\delta'$---a magnitude cannot be
made odd without becoming a cardinality difference (\S\ref{sec:symm}).
The notation needs a
word. $B\triangle S$ is the \emph{symmetric difference} of the two sets: the
neighbour counts at which birth and survival \emph{disagree}---a count belongs to
$B\triangle S$ when it triggers exactly one of the two, birth or survival, but not
both and not neither. For Conway ($B=\{3\}$, $S=\{2,3\}$) the two sets share the
count $3$ and differ only at $2$, so $B\triangle S=\{2\}$: a cell's own state
matters only when it has exactly two live neighbours. Two consequences follow.
\begin{enumerate}
  \item \textbf{$\mu_{\text{centre}}=0$ is the $B=S$ submanifold.} By the formula
    $\mu_{\text{centre}}=0$ iff $B\triangle S=\varnothing$, i.e.\ $B=S$. So
    $\mu_{\text{centre}}$ is exactly the binomially-weighted set distance from
    centre-blind, and the $512$ centre-blind rules are characterised a priori:
    they cannot respond differently to the centre cell's own state at any count.
  \item \textbf{$\mu$ is independent of $\lambda'$}: one measures the fraction of
    table entries that fire, the other the fraction of input-bit flips that
    change the output. Together they form a Binder $(\lambda',\mu)$ plane
    (\S\ref{sec:results-mu}).
\end{enumerate}

\subsection{A symmetry-organised coordinate system}
\label{sec:symm}

The rule table carries two natural involutions,
\begin{equation}
  \tau:(B,S)\mapsto(S,B),\qquad
  \sigma:(B,S)\mapsto\bigl(\overline{\mathrm{rev}(S)},\overline{\mathrm{rev}(B)}\bigr),
\end{equation}
where $\mathrm{rev}(M)=\{8-k:k\in M\}$ and the bar is the complement in
$\{0,\dots,8\}$. $\tau$ swaps birth and survival; $\sigma$ is the bit-flip
duality, in the sense that if $g=\sigma(f)$ then $g(\lnot x)=\lnot f(x)$ for every
configuration. Both square to the identity, and the rule space falls into the
four parity sectors of $(\tau,\sigma)$. Of these four sectors, three are
populated by coordinates the paper uses: $\lambda'$ occupies
($\tau$-even, $\sigma$-odd), $\delta'$ occupies ($\tau$-odd, $\sigma$-even),
and $\mu$ (with $\mu_{\text{centre}}$) occupies ($\tau$-even, $\sigma$-even).
The remaining sector, ($\tau$-odd, $\sigma$-odd), is not empty in
principle---a natural candidate is an edge-contiguity coordinate that
compares how bunched the two thresholds are as bands of neighbour
counts---but the one such coordinate I constructed and tested proved
subordinate to $\delta'$: conditional on $(\lambda',\delta',\mu_{\text{centre}})$
it retained only a small fraction of $\delta'$'s effect on $\phi>0$, so I do
not carry it through the analysis (\S\ref{sec:disc-limits}).

\paragraph{Parity table.}
I verified the action of $\tau$ and $\sigma$ on every coordinate by direct
evaluation on all $262{,}144$ rules (max residue $1.1\times10^{-16}$): $\lambda'$
is ($\tau$-even, $\sigma$-odd; $\lambda'\mapsto1-\lambda'$); $\delta'$ is
($\tau$-odd, $\sigma$-even; $\delta'$ invariant under $\sigma$);
$\mu_{\text{centre}},\mu$ are ($\tau$-even, $\sigma$-even).
The key observation is that $\lambda'$ and $\mu$ are both
$\tau$-even, so neither can distinguish a rule from its $B\leftrightarrow S$
mirror; $\delta'$ is the unique $\tau$-odd coordinate among the
activity/sensitivity axes, and is therefore the natural birth-versus-survival
scalar. Since $\sigma$ acts as
$(\lambda',\delta',\mu)\mapsto(1-\lambda',\delta',\mu)$, self-dual rules need
not have $\delta'=0$---consistent with Day~\&~Night ($\delta'=-\tfrac19$)
being self-dual. Pearson correlations across the full rule space are exactly
zero between every pair of $\{\lambda',\delta',\mu\}$, as required by the
orthogonality of parity sectors under the uniform measure.

\paragraph{The table is forced, not chosen.}
That the sensitivity coordinates all occupy one sector is not an artefact of
which observables I happened to compute. A Boolean derivative asks only whether
a single-bit flip changes the output; it is a magnitude, carrying no sign and no
direction, whereas an odd coordinate must carry one. Concretely, if
$g=\sigma(f)$ then $g(\neg x)=\neg f(x)$ for every configuration, so flipping
input bit $i$ changes $g$'s output at $\neg x$ exactly when it changes $f$'s at
$x$; since $x\mapsto\neg x$ preserves the uniform measure, $\mu(g)=\mu(f)$.
Likewise $\tau$ leaves $\mathbf 1[n\in B\triangle S]$ fixed and symmetrises the
neighbour term. \emph{Every} sensitivity of this kind is therefore ($\tau$-even,
$\sigma$-even), and the remaining three sectors can be populated only by
counting functionals of the rule table; the two families are not interchangeable
options for a cell. A second constraint fixes which coordinate can carry the
exclusion. Any $\tau$-odd $f$ obeys $f(\tau x)=-f(x)$, so it vanishes on the
fixed-point set $\{B=S\}$ of $\tau$---but also wherever its two sides merely
balance, a strictly larger set ($\delta'=0$ on $48{,}620$ rules). Only a
$\tau$-even functional that is \emph{non-separable}, a function of $B\triangle S$
jointly rather than of $B$ and $S$ apart, can have $\{B=S\}$ as its exact zero
set. $\mu_{\text{centre}}$ is such a functional, which is why the exclusion of
\S\ref{sec:results-sigma} is stated with it and could not have been stated with
$\delta'$; its binomial weighting is inherited from the configuration measure and
plays no part, since any strictly positive weighting of $B\triangle S$ has the
same zero set. The exclusion depends on the disagreement set being empty, not on
how the disagreement is weighted.

\paragraph{Roles of the coordinates.}
Each answers a distinct question the others structurally cannot, and the
necessary/sufficient distinction runs through all three. $\lambda'$ places a rule on
the frozen$\to$chaotic spectrum but collapses rules of equal total firing
fraction, so it is a necessary entry point, never sufficient. $\delta'$ is the
swap-odd factorisation axis: $\phi>0$ rules are birth-light and survival-rich
($\delta'\approx-0.22$), the rule-table fingerprint of the $B$-gate / $S$-knob
structure (\S\ref{sec:results-factor}); without $\delta'$ the claim cannot even be
stated. $\mu_{\text{centre}}$ is the one coordinate whose zero set coincides
exactly with the $B=S$ diagonal, so it labels the $B=S$ exclusion in a
single coordinate value; this is a re-parameterisation of the exclusion, not
a mechanism for it (\S\ref{sec:results-sigma}, \S\ref{sec:disc-mechanism}).

\subsection{Combinatorial enumeration: $\phi$, $\psi$, $\omega$}
\label{sec:enum}

The following are functions of the rule table alone and independent of any
simulation protocol. For each rule I enumerate every $3{\times}3$ binary pattern
(reduced to its $D_4$ orbit representative) on a quiescent zero-padded background,
evolve it to $T_{\max}=16$, and classify the orbit by its return behaviour: a
nonzero-offset translated return is a \emph{spaceship} ($\phi$ counts distinct
orbit-representative spaceships, with period and velocity set); a same-position
return at period $2\le T\le16$ is an \emph{oscillator} ($\omega$); a fixed pattern
($T{=}1$) is a \emph{still life} ($\psi$). ``$3{\times}3$'' refers to the
\emph{seed footprint}: the pattern may expand during the cycle and must contract
back at $t=T$; $\phi$ therefore counts compact-\emph{seed} spaceships of period
$\le 16$, not spaceships whose bounding box stays inside $3{\times}3$ throughout.
The period cut-off is an empirical truncation but is
close to saturating: the min-period distribution of the $424$ $\phi>0$ rules
peaks sharply at $T=4$ ($312$) with a rapidly decaying tail ($T=5{:}48$,
$T=6{:}2$, $T=7{:}48$, $T=8{:}8$, $T=10{:}4$, $T=13{:}1$, $T=14{:}1$, none at $T=15$ or
$T=16$), so the cutoff censors negligibly. Directed spaceship searches over
life-like rules exist~\cite{eppstein2002}; my enumeration differs in being
exhaustive over all $D_4$-orbit seeds of a fixed footprint and all $262{,}144$
rules simultaneously. $\phi$ asks ``can the rule
support a compact glider at all,'' independent of whether a random trajectory ever
exhibits one.

\paragraph{Protocol specifics.} Each seed is placed centred on a padded grid
of size $2T_{\max}\cdot v_{\max}+n+2$ (in cells, where $n=3$ is the seed
footprint, $v_{\max}=1$ cell/step is the maximum life-like propagation speed,
and the extra padding guarantees no live cell reaches the boundary within
$T_{\max}=16$ steps). Evolution halts if the pattern extincts, if the
bounding box exceeds $9{\times}9$ (a growth escape criterion; the seed is
then not classified), or at $t=T_{\max}$. Return classification checks
whether the current live set is a rigid translate of the seed's live set
(with any integer offset $(dx,dy)\ne(0,0)$): I do not accept glide-reflected
returns as spaceships at this footprint, which is a convention choice, not a
mathematical necessity---glide-reflected returns at compact footprint are
possible in principle for some rules but were not observed under this
protocol. The pre-screen relies on a necessary condition: if a phase at
$t\in\{1,\ldots,T_{\max}\}$ is a translate of the seed, both configurations
have the same live-cell count. I evaluate the count at every step and short-
circuit the full shape-comparison only when the count matches the seed's
count; this is a check applied at the potential return step, not a
mid-cycle rejection (mid-cycle populations can differ from the seed's, and
the pre-screen never prunes them). No false negatives: any true translated
return is caught because the count equality is a necessary condition, and
the shape comparison at return steps is exhaustive.

\subsection{Dynamical probe and coherent-propagation index $\mathcal{G}^{*}$}
\label{sec:dyn}

To identify rules whose combinatorial propagation is or is not expressed by random
dynamics I run a single fixed sweep. For each rule I draw $5$ shared Bernoulli
ICs at density $\rho_0=0.30$ on a $128\times128$ torus, together with their
bit-flipped partners at density $0.70$ (the symmetric-IC ensemble described in
the next paragraph), evolve $600$ steps, and measure on the $200$-frame
tail a residual spatiotemporal autocorrelation $\mathcal{G}^{*}_{\text{alt}}$: a
coherent structure of period $T$ and velocity $(v_x,v_y)$ produces an off-axis
peak at $(\Delta t,\Delta x,\Delta y)=(T,-Tv_x,-Tv_y)$, validated by an injection
test in which an isolated Conway glider peaks at $(\Delta t{=}4,\Delta
x{=}{-}1,\Delta y{=}{-}1)$. I use $\mathcal{G}^{*}_{\text{alt}}$ only to flag
rules with no measurable coherent propagation; the remaining dynamical
observables of the sweep are regime descriptors, reported as such.

\paragraph{Bit-flip duality and the symmetric-IC ensemble.}
The bit-flip duality $\sigma$ (\S\ref{sec:symm}) is exact at the level of
trajectories ($g(\lnot x)=\lnot f(x)$ for $g=\sigma(f)$) but is broken by sparse
Bernoulli ICs: a rule and its dual visit different attractors from the same IC.
Pairing every sample with its bit-flipped twin at density $1-\rho_0$ restores
the duality at the level of ensemble means; I verify a residual twin-gap at
the seed-noise floor. The $512$ $\sigma$-fixed points (e.g.\ Day~\&~Night) are
the self-dual rules. Beyond $\mathcal{G}^{*}_{\text{alt}}$ the sweep records
per-frame compressibility, a basin-asymmetry index and the cell-flip rate;
gauge-fixed by symmetric-IC averaging these co-locate $\sigma$-twins and
separate canonical anchors into recognisable regimes. They are reported
primarily as regime descriptors, deferred to the repository; capability
detection through them is a partial secondary signal rather than the primary
route.

\section{Results}

\subsection{The rule-table phase diagrams $(\lambda',\mu)$, $(\delta',\mu)$, $(\lambda',\delta')$}
\label{sec:results-mu}

\begin{figure}[p]
\centering
\includegraphics[width=0.715\linewidth,keepaspectratio]{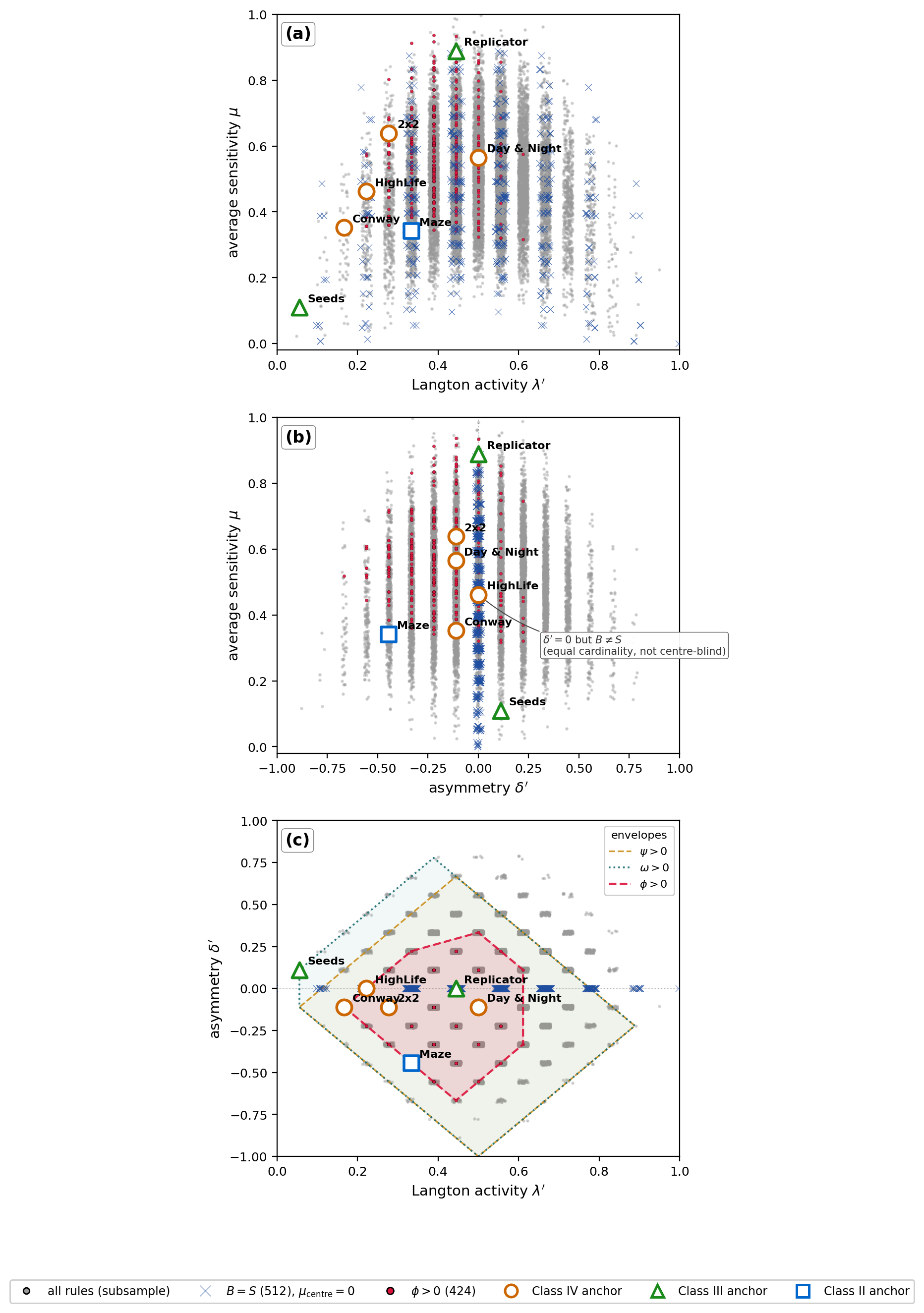}
\caption{Coordinate projections of all $262{,}144$ life-like rules. Grey:
subsample of the full rule space. Royal-blue X: the $512$ centre-blind
($B=S$) rules ($\mu_{\text{centre}}=0$). Crimson: the $424$ rules with
$\phi>0$. Anchors are outlined markers coloured by Wolfram class.
(a)~$(\lambda',\mu)$: $B=S$ rules trace the $\mu_{\text{centre}}=0$ locus.
(b)~$(\delta',\mu)$: $B=S$ rules collapse to $\delta'=0$ while $\phi>0$
rules cluster at $\delta'\approx-0.2$; HighLife lies on $\delta'=0$ by equal
cardinality without being centre-blind (§\ref{sec:coords}).
(c)~$(\lambda',\delta')$: three nested convex-hull envelopes show the
propagation-storage-oscillation factorisation of §\ref{sec:results-factor}
in coordinate space---crimson dashed for $\phi>0$ ($n=424$, 6 vertices),
teal dotted for $\omega>0$ ($n=57{,}270$), amber dashed for $\psi>0$
($n=62{,}548$). Envelopes are geometric necessary regions but not
coordinate-level exclusions: the crimson envelope alone contains
$\sim\!204{,}560$ $\phi=0$ rules ($\sim\!78\%$ of the $\phi=0$ bulk), and
the amber envelope covers most of the attainable plane.}
\label{fig:planes}
\end{figure}

The sensitivity coordinate places every rule on a second axis complementing
$\lambda'$ (Figure~\ref{fig:planes}). The cloud of all $262{,}144$ rules forms a
structured band, and the $B=S$ submanifold ($\mu_{\text{centre}}=0$) traces a
sharply identifiable locus on which all $512$ centre-blind rules lie.
Replicator sits at the top with $\mu=0.889$---centre-insensitive but maximally
neighbour-sensitive---showing $B=S$ is not synonymous with low sensitivity, only
with centre-flip insensitivity. The $424$ rules with $\phi>0$ cluster off that
locus at moderate $\lambda'$ and moderate $\mu$: full range
$[0.32,0.94]$, interquartile range $[0.44,0.71]$, median $0.55$. Conway
($\mu=0.35$) sits at the low-$\mu$ edge of the $\phi>0$ cluster and HighLife
($0.46$) inside it, while the storage-heavy Class~IV anchor Day~\&~Night
lies at $\mu=0.565$ yet is itself $\phi=0$ (Figure~\ref{fig:joint})---a
first illustration that inhabiting the $\phi>0$ region on $(\lambda',\mu)$
is necessary for compact propagation but does not certify
it~\cite{wolfram1984}. ($2{\times}2$, another Class~IV anchor with
$\phi=0$, sits at $\mu=0.639$, inside the $\phi>0$ IQR; both rules are
complex through engineered rather than compact-glider dynamics, a point
I return to in \S\ref{sec:disc-conway}.) The middle panel shows the
same rules in $(\delta',\mu)$, where the $\phi>0$ cluster sits at
$\delta'\approx-0.2$ (the birth-light/survival-rich offset of
\S\ref{sec:results-factor}) and the $B=S$ rules collapse to the line $\delta'=0$;
the right panel shows the $(\lambda',\delta')$ projection, where the $\phi>0$
rules form a diamond near $(0.4,-0.2)$. I do not claim the planes as a sharp
classifier: the $\phi>0$ band is contained in, but does not exhaust, the
moderate region---necessary, not sufficient.

\subsection{Combinatorial spaceship support and the $B=S$ exclusion}
\label{sec:results-sigma}

Of the $262{,}144$ rules, $\phi>0$ holds for $424$ ($0.16\%$); these cluster in
the $(\lambda',\delta')$ diamond centred near $(0.4,-0.2)$
(Figure~\ref{fig:planes}c, innermost crimson envelope), with Conway at
its low-$\lambda'$ edge. They support
$876$ distinct orbit-representative spaceships, of which $560$ are diagonal and
$316$ orthogonal, with no $3{\times}3$ knight-ship (an empirical finding of
the exhaustive enumeration, not a geometric impossibility: a knight-vector
translation is compatible with a $3{\times}3$ seed and $T\le 16$ in principle,
but no rule realises one).
The period distribution peaks at $T=4$. Direction is a clean per-rule property: of the $424$, $300$ support only
diagonal ships and $124$ only orthogonal, none both at the $3{\times}3$ footprint.
Only Conway and HighLife appear among the canonical anchors, leaving many
rules that support a compact glider yet remain uncatalogued in prior
searches~\cite{eppstein2002,goucher_catagolue}---though, as
\S\ref{sec:disc-conway} shows, those with genuinely Conway-like bulk dynamics turn
out to be anchor variants rather than new rules.

\paragraph{The $B=S$ exclusion and its footprint dependence.}
The $512$ rules with $B=S$ (equivalently $\mu_{\text{centre}}=0$) contain
\emph{zero} rules with $\phi>0$ at any $T\le16$, against the bulk rate of
$0.16\%$. This is the sharp form of ``exclude, not certify'': the $B=S$ diagonal
(equivalently the $\mu_{\text{centre}}=0$ level set) is a coordinate-level
label for the empirical exclusion. It is specific to the compact
footprint. Enumerating larger seeds exhaustively over all $B=S$ rules, $36$
of the $256$ vacuum-stable ones support a $4{\times}4$ spaceship and $40$ a
$5{\times}5$ one (against $0$ at $3{\times}3$), so the exclusion holds at
the compact footprint alone, not at all scales. ($\phi$ enumerates exactly that
compact channel, not ordinary activity spread, which any non-frozen rule
supports.) The
recovery is bounded on the two footprints examined: the count sits near
$15\%$ of vacuum-stable $B=S$ rules at both $4\times 4$ and $5\times 5$
(a two-point observation, not extrapolable), and \emph{build-and-hold}---a glider together with
static storage---remains rare, holding for $4$ rules at $4{\times}4$ and $8$ at
$5{\times}5$ (counting still lifes; every glider-bearing rule oscillates, so
oscillation is not an informative storage test here). Moreover the gap to Conway
\emph{widens} with footprint: the still-life count of the richest such $B=S$ rule
trails Conway's by $6.3\times$ at $4{\times}4$ and $9.2\times$ at $5{\times}5$. A
rule forced onto the diagonal tunes propagation and storage with a single set and
cannot exploit a larger window on both axes at once, as a rule with independent
$B$ and $S$ can; the strict $3{\times}3$ exclusion therefore softens into a
widening handicap rather than vanishing. The combinatorial recovery does not,
however, bring behavioural complexity with it. Running the richest build-and-hold
candidates---the rules that combine a larger-footprint glider with the most still
lifes---from random initial conditions across a range of densities, and inspecting
the trajectories directly, I find chaotic (Class~III-like) bulk dynamics: the
field boils to a disordered steady state without the enumerated glider visibly
emerging or persistent objects surviving. Within the footprints examined
($\le5{\times}5$), no centre-blind rule was found to express its enumerated glider
from a random start: the diagonal gains combinatorial glider capability at larger
scales without expressing it. A broader
behavioural screen, applied without reference to $\phi$, flags a small number of
centre-blind rules as candidates for structured dynamics; whether these constitute a
genuine small-object ecology or densely-packed frozen tessellations is not settled
by the sampling used here, and I leave it open. Discussion of possible
partial mechanisms and their limits is in \S\ref{sec:disc-mechanism}.
(The $4{\times}4$ enumeration is exhaustive over all
$262{,}144$ rules---$8{,}206$ support a spaceship, and the $424$ compact-glider
rules are contained in this set exactly; the $5{\times}5$ enumeration is
exhaustive over $B=S$ but sampled elsewhere. Enumeration at $6{\times}6$ and
larger footprints is beyond my computational means, so I cannot rule out that
the $B=S$ recovery continues to widen or eventually saturates further out.)

\begin{figure}[t]
\centering
\includegraphics[width=\linewidth]{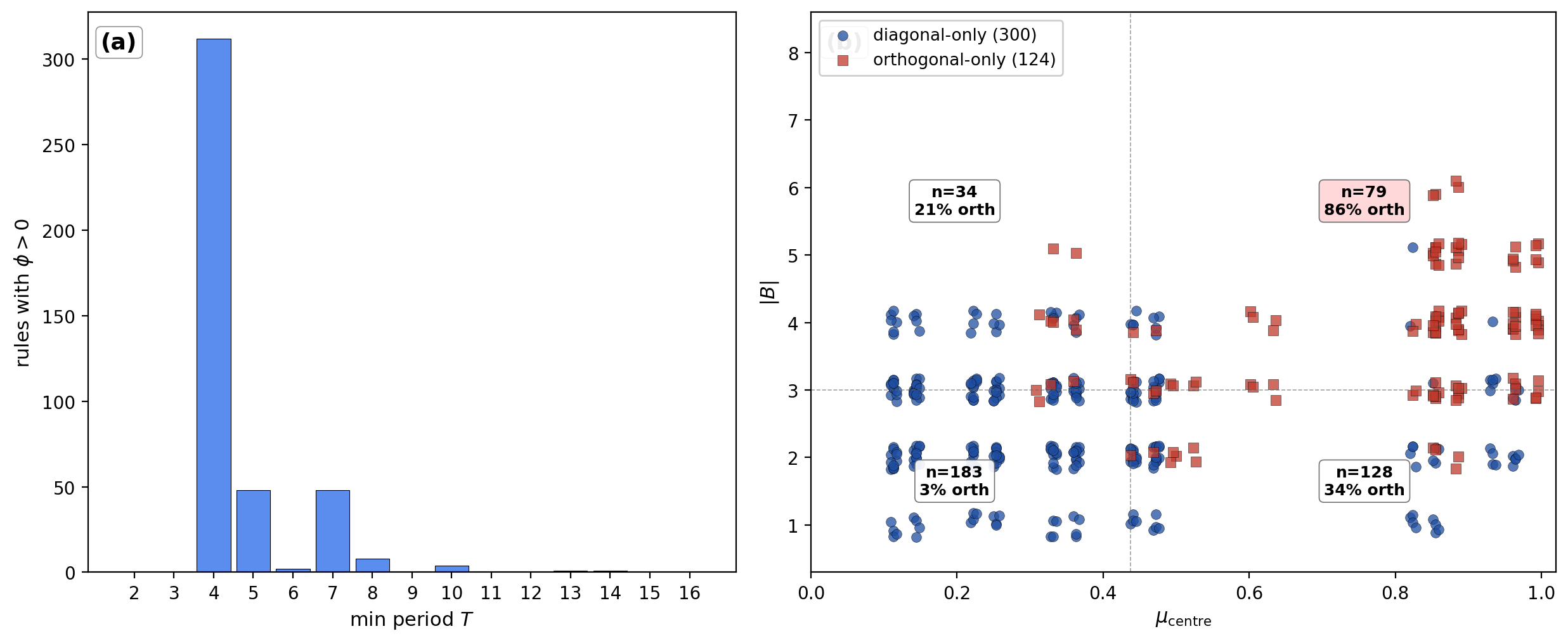}
\caption{Combinatorial characterisation of the $424$ rules with $\phi>0$.
(a)~Period distribution, mode $T=4$; the tail decays rapidly with a single
rule at $T=13$ and a single rule at $T=14$, none at $T=15$ or $T=16$, so the
enumeration's $T\le 16$ cutoff censors negligibly.
(b)~Direction in the $(\mu_{\text{centre}},|B|)$ plane, points coloured by
whether the rule supports only diagonal ships (blue circles) or only
orthogonal ships (red squares); no rule at the compact footprint supports
both. Orthogonal motion concentrates in high $\mu_{\text{centre}}$ \emph{and}
large $|B|$ jointly---the joint (upper-right) corner is $86\%$ orthogonal,
either lever alone $21$--$34\%$, and the unfavoured (lower-left) corner
$3\%$; diagonal is the default elsewhere
(\S\ref{sec:disc-mechanism}).}
\label{fig:sigma}
\end{figure}

\subsection{Glider-capable but dynamically silent rules}
\label{sec:results-silent}

Cross-referencing $\phi$ against $\mathcal{G}^{*}_{\text{alt}}$ isolates a sharp
population that makes the capability/expression gap concrete. I find $33$ rules
that combinatorially support a $3{\times}3$ glider yet produce no coherent
propagation from random initial conditions across a deliberately wide search---$13$
densities $\rho_0\in[0.05,0.90]$, $25$ seeds each, runs of $10^{3}$ steps ($325$
trials per rule). The asymmetry between the two claims is the point. That the
glider \emph{exists} is a construction, not a sample: enumerated from the table and
verified by injecting it onto a quiescent grid, where it propagates steadily. That
random dynamics \emph{do not} produce it is, by contrast, only ever an estimate
over sampled initial conditions; I state it as a measure, not an absolute---the
glider's basin is negligible under random sampling, undetected in those $325$
trials. Capability is certifiable from the table; expression is only estimable
from dynamics. (A further five rules silent at $\rho_0=0.30$ express intermittently
at other densities; they are rare-basin, not silent, and are excluded from the
$33$.)

The obstruction is not a missing background. Every $\phi>0$ rule has a stable
dead vacuum by construction---a glider is enumerated on a quiescent background, so
a rule whose vacuum ignited could not register $\phi>0$---and indeed all $33$
have $0\notin B$. The glider genuinely lives on empty space; the random-IC
trajectory simply never assembles it there. Measuring the trajectories shows two
opposite failures: $22$ of the $33$ \emph{freeze} (the field settles to
near-vacuum, observed tail densities $\bar\rho\approx0.005$--$0.012$) and $11$
\emph{flood} (the field saturates, $\bar\rho\approx0.58$--$0.99$), with none in
the moderate band $[0.10,0.56]$ that expressing rules---Conway among them---occupy.
On the silent population the split is read by a single survival bit: every flood
rule has $8\in S$ (saturated regions persist; $11/11$), while the freeze rules
have restrictive survival (mean $|S|\approx4.5$ across the freeze rules,
two-thirds with loneliness death
$0,1\notin S$). This is descriptive of the $33$, not a general law: on a uniform
$3{,}000$-rule sample from the full $262{,}144$, ranking rules by activity
$\lambda'$ alone recovers the flood-versus-freeze (dense-versus-sparse
settling) split in the large majority of pairwise comparisons---about
$86\%$---while no simple birth/survival set relation predicts the split
out of sample. The silent rules are those
whose glider-bearing birth set pairs with a survival rule that drives the
trajectory out of the viable band, too dense or too sparse for the glider to
persist on clean background. Which regime a rule falls into is determined by the
table but not reducible to a closed-form predictor of it: a mean-field density
map in the style of Schulman and
Seiden~(1978)\nocite{schulmanseiden1978}
reproduces the gross density ordering across rules (rank correlation $0.77$ with
measured $\bar\rho$, against $0.34$ for the best single-bit feature) yet
mis-places most silent rules at a moderate interior fixed point
($\rho\approx0.4$) the spatially-fragmenting dynamics never reach. The regime must
be simulated---the dynamical counterpart of ``certifiable versus estimable.''

A controlled witness fixes capability and isolates the survival rule. The family
$B=\{3\}$ contains rules that share a common supported diagonal glider, Conway
among them. Holding the glider fixed and varying only $S$, some express it from
random ICs and others do not: Conway's $S=\{2,3\}$ does, while
$B\{3\}/S\{2,4,5,8\}$---the same birth set, the same supported glider---does not.
Expression is thus not a function of capability alone. Propagation capacity is set
by $B$; whether random dynamics express it depends on $S$ (\S\ref{sec:disc-conway}).

\subsection{Hierarchical $B/S$ factorisation of propagation and storage}
\label{sec:results-factor}

\begin{figure}[t]
\centering
\includegraphics[width=0.92\linewidth]{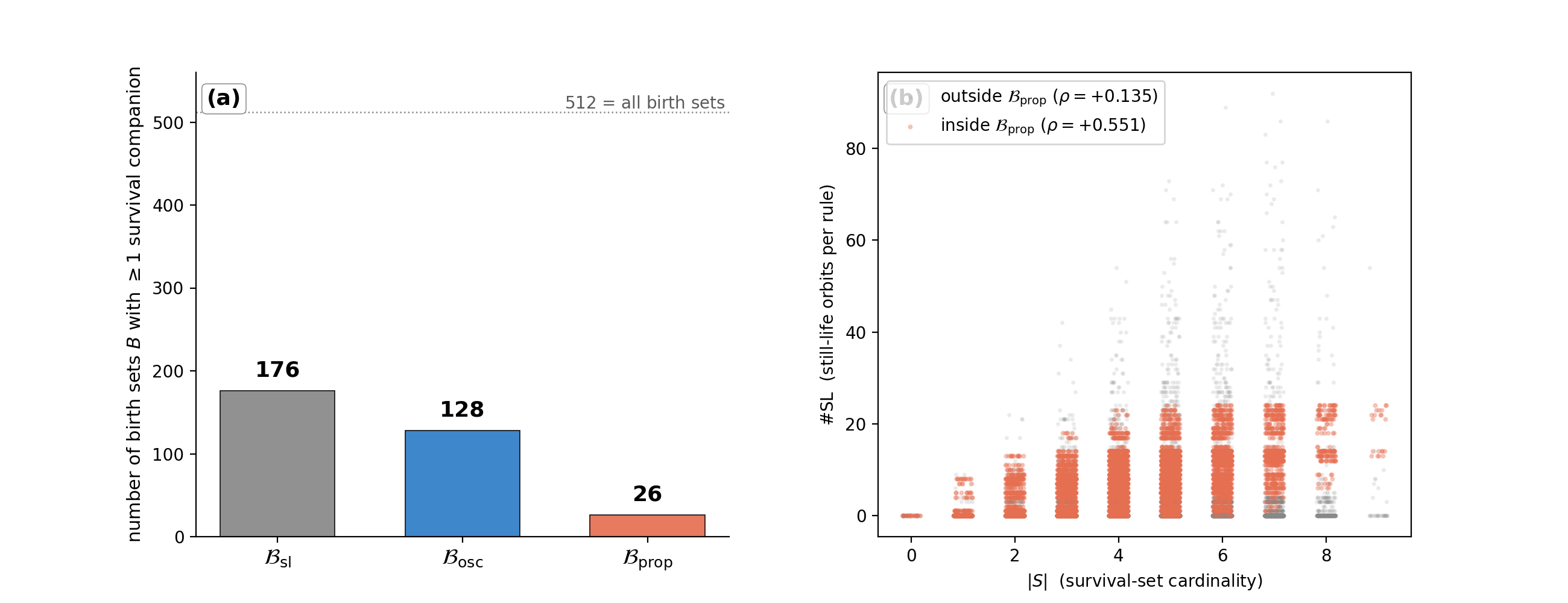}
\caption{Hierarchical $B/S$ factorisation.
(a)~Birth-set support of each capability (a birth set qualifies iff at least
one of the $512$ possible survival companions $S$ yields the structure, i.e.\
$k\ge 1$):
$\mathcal{B}_{\text{sl}}\supset\mathcal{B}_{\text{osc}}\supset\mathcal{B}_{\text{prop}}$
with $176$ (still lifes), $128$ (oscillators) and $26$ (spaceships) birth
sets respectively, nested strictly within the $512$ possible---propagation is
the narrowest gate, oscillation wider, storage widest. (b)~$\#\mathrm{SL}$
against survival-set cardinality $|S|$, split by $B\in\mathcal{B}_{\text{prop}}$
(orange) vs.\ outside (grey): inside the propagation gate storage tunes with
$|S|$ (Spearman $\rho=0.551$), outside it does not ($\rho=0.135$).}
\label{fig:factor}
\end{figure}

Define the propagation-enabling birth sets $\mathcal{B}_{\text{prop}}=\{B:|\{S:
\phi(B,S)>0\}|\ge1\}$. Only $26$ of $512$ birth sets ($5.1\%$) qualify, and
by the reduction of \S\ref{sec:disc-mechanism} every one of them contains
$3$ (and none contains $2$); outside $\mathcal{B}_{\text{prop}}$
propagation capacity is absent by construction. Within the enabled slice ($13{,}312$
rules) the Spearman correlation between still-life count and $|S|$ is $0.551$;
outside the slice it drops to $0.135$ (Figure~\ref{fig:factor}). The factorisation
is hierarchical, not symmetric: $B$ gates whether any structure is possible, $S$
tunes storage richness once the gate is open. One caveat on what is being
discovered here: that a still life's persistence involves $B$ (empty
neighbourhoods of the still life must not birth) is partly definitional --- a
$3{\times}3$ enumeration on a quiescent background necessarily reads any
would-be still life against the birth condition. What is not definitional, and
is what the enumeration actually finds, is the \emph{strict} nesting
$\mathcal{B}_{\text{prop}}\subset\mathcal{B}_{\text{osc}}\subset\mathcal{B}_{\text{sl}}$
with the specific ratios $26/128/176$ (out of $512$) and the four-fold Spearman
contrast $0.551$ vs.\ $0.135$; both quantify how much $S$-tuning happens
\emph{conditional on} the $B$-gate.

\paragraph{In-place recurrence forms one storage axis.}
Enumerating in-place oscillators by the same protocol, $57{,}270$ rules
($21.85\%$) have $\omega>0$; all $424$ $\phi>0$ rules are among them, $56{,}846$
oscillate without any spaceship, and none support a spaceship without also
oscillating. Oscillation is birth-gated through a strictly wider gate:
$\mathcal{B}_{\text{osc}}$ contains $128$ birth sets against $26$ in
$\mathcal{B}_{\text{prop}}$, and storage (any still life) through a wider one
still, $176$ in $\mathcal{B}_{\text{sl}}$; the three form a strict three-level
nesting $\mathcal{B}_{\text{prop}}\subset\mathcal{B}_{\text{osc}}
\subset\mathcal{B}_{\text{sl}}$ ($26\subset128\subset176$ of $512$;
Figure~\ref{fig:factor}a).
The same nesting appears in coordinate space as three nested convex-hull
envelopes on the $(\lambda',\delta')$ plane (Figure~\ref{fig:planes}c). The
coordinate-space view understates the sharpness of the factorisation: the
amber storage envelope covers most of the attainable plane because $\psi>0$
holds for $\sim\!24\%$ of rules, so ``inside the storage envelope'' is a
very loose necessary condition on $(\lambda',\delta')$. The tight version of
the factorisation---$B\in\mathcal{B}_{\text{sl}}$ for storage,
$B\in\mathcal{B}_{\text{osc}}$ for oscillation,
$B\in\mathcal{B}_{\text{prop}}$ for propagation---is a birth-set statement
(Figure~\ref{fig:factor}a), not a cardinality one; the cardinality
coordinates locate but do not certify, and the visible gap between the three
envelopes on Figure~\ref{fig:planes}c is a coarse silhouette of a much
sharper birth-set structure.
Storage is thus wider than oscillation---modestly, not vastly---and
propagation is the narrowest gate. I find
Spearman $\rho(\omega,\psi)=+0.825$ against $\rho(\omega,\phi)=+0.080$;
the informative quantity is the contrast, not the magnitude (still lifes and
oscillators are the same enumeration at period $1$ and $\ge2$, so a strong
correlation is expected). The relevant partition is not static-versus-moving but
translating-versus-non-translating: in-place recurrence at any period forms a
single $S$-tuned storage axis, and translation is empirically separated
from it at $3{\times}3$: only rules with $\mu_{\text{centre}}>0$ (i.e.\
$B\neq S$) show compact-glider support, while in-place recurrence occurs on
both sides of the $\mu_{\text{centre}}=0$ line. The $B=S$ submanifold
confirms this directly: $B=S$
rules support oscillators at $96/512=18.75\%$ (within reach of the $21.85\%$
full-space rate, all at period $T=2$) while supporting $0/512$ compact spaceships.
$\mu_{\text{centre}}=0$ closes the translation channel specifically and leaves the
in-place storage channel essentially intact (Figure~\ref{fig:factor}; mechanism in
\S\ref{sec:disc-mechanism}).

\begin{figure}[t]
\centering
\includegraphics[width=0.72\linewidth]{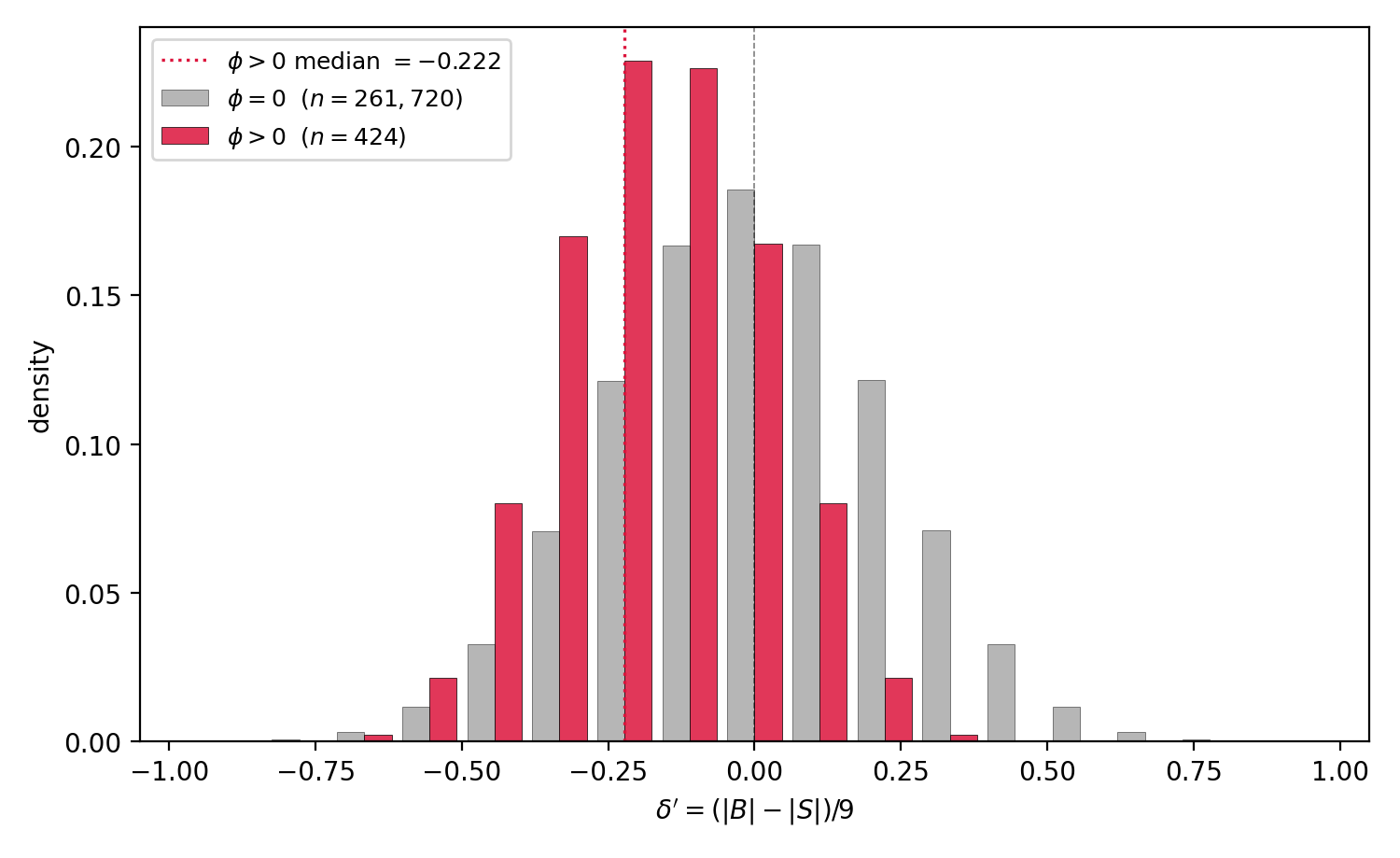}
\caption{Cardinality asymmetry $\delta'=(|B|-|S|)/9$ over all $262{,}144$
rules. Grey: $\phi=0$ bulk, symmetric about zero (forced by
$\tau$-symmetry). Crimson: $\phi>0$ ($n=424$), sharply skewed to
$\delta'<0$ with median $-0.222$; $340$ of the $424$ compact-glider rules
sit at $\delta'<0$.}
\label{fig:tauodd}
\end{figure}

\paragraph{$\delta'$ as the factorisation axis.}
Because the factorisation is intrinsically a birth-versus-survival statement, the
$\tau$-odd coordinate $\delta'$ is its natural projection. Among the $424$
$\phi>0$ rules the $\delta'$ distribution is sharply skewed to $\delta'<0$ (median
$-0.222$, $340$ of $424$ below zero), whereas the $\phi=0$ bulk is symmetric
about zero by the $\tau$-symmetry of rule space (Figure~\ref{fig:tauodd}). This
$\delta'\approx-0.22$ plays, for the asymmetry axis, the role Langton's critical
value played for the activity axis---a single scalar where capability
concentrates---though the comparison is qualitative: Langton's $\lambda\approx0.25$
is on his multiplicity-weighted scale, not directly comparable to my unweighted
$\lambda'$ (\S\ref{sec:coords}). But the paper's framing applies to it in
full---both $\lambda'$ and $\delta'$ \emph{locate} without \emph{certifying}. The same
$\delta'\approx-0.22$ slice is shared by thousands of $\phi=0$ rules, so landing
on it is necessary for the propagation--storage structure but never sufficient for
a glider; the value marks a region, not a guarantee. This fingerprint is specific
to the compact footprint: among the $8{,}206$ rules that support a $4{\times}4$
spaceship the $\delta'$ skew washes out (median $0$, rank-biserial $-0.12$ against
$-0.42$ at $3{\times}3$), because a wider window admits birth-heavier
orthogonal shippers. What survives footprint widening is the \emph{gate}
structure, not the $\delta'$ sign: the $26$ birth sets of
$\mathcal{B}_{\text{prop}}^{3\times3}$ are contained in the $64$ of
$\mathcal{B}_{\text{prop}}^{4\times4}$ exactly (both counted at the
existence threshold $k\ge1$), and both remain dominated by small
birth sets (mean $|B|=3.5$ at $4{\times}4$, against a uniform peak at $|B|=5$). The
birth-light asymmetry is thus what \emph{compact} propagation costs, not a
universal law; the narrow gate of small birth sets is the footprint-robust
core. $\delta'$ is the only odd coordinate the results here require;
auxiliary odd-sector coordinates were explored but do not carry additional
capability information beyond what $\delta'$ already provides
(\S\ref{sec:disc-limits}).

\begin{figure}[t]
\centering
\includegraphics[width=0.7\linewidth]{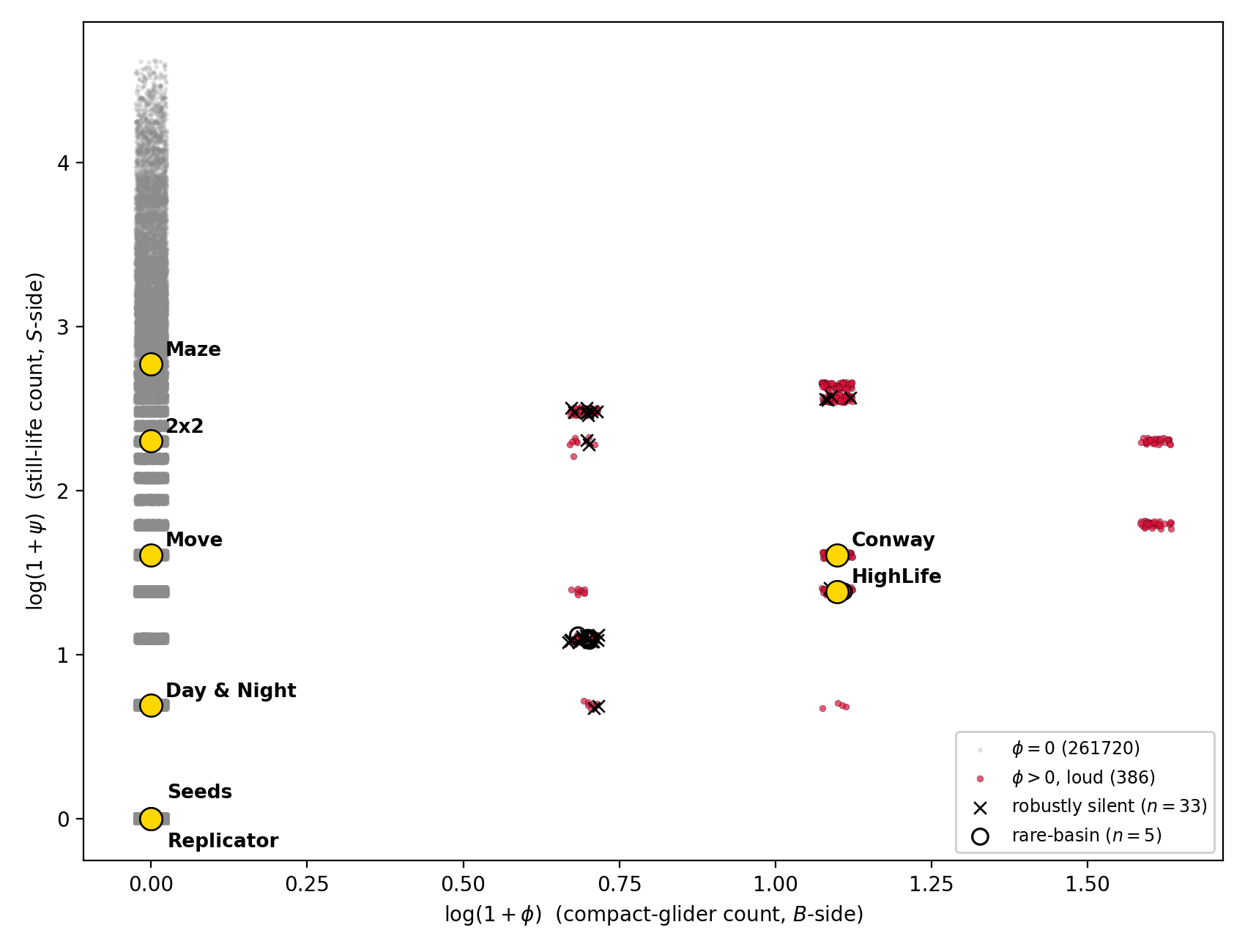}
\caption{Joint propagation--storage plane $(\phi,\psi)$ in $\log(1+x)$
coordinates. Conway and HighLife sit in the upper-right corner (both gliders and
rich storage). The $\phi=0$ column supports no \emph{compact} glider and
contains the Class~IV anchors $2{\times}2$, Day~\&~Night and Move---whose
complexity rests on engineered, large-footprint propagation this enumeration does
not capture (\S\ref{sec:disc-conway})---while Seeds and Replicator sit at the
origin. The silent population from the main sweep is shown split into $33$
robustly silent rules (crosses) and $5$ rare-basin rules (open circles), per
\S\ref{sec:results-silent}; the two together total $38$ points.}
\label{fig:joint}
\end{figure}

\section{Discussion}

\subsection{The mechanism of the $B=S$ exclusion}
\label{sec:disc-mechanism}

``Propagation'' here means coherent translation: a pattern returning to its own
configuration displaced by a fixed offset, keeping its shape as it moves.
It is not activity spread---disturbances travelling outward at up to one cell
per step through a churning background, which any non-frozen rule does, $B=S$
rules included. The exclusion below concerns coherent movers, not motion.

A compact spaceship must create state ahead of itself and clear state behind,
both enacted by the same object in the same update. What changes with footprint
is how many vantage points the object provides for carrying that difference. A
$3{\times}3$ pattern is exactly the Moore neighbourhood of a single cell: one
cell's update depends on the whole object, and no other cell---inside or
outside---sees all of it. Indeed $3{\times}3$ is the largest pattern contained
in a single neighbourhood, which is what makes ``compact'' intrinsic rather
than an arbitrary window. At that footprint the front/back distinction must
pass through one centre, and $\mu_{\text{centre}}=0$ makes the update a
function of the neighbour count alone. This is the heuristic; the argument
that follows makes it concrete by identifying the specific rule-table
resource a compact glider needs and $B=S$ closes.

\paragraph{State-distinction channels.}
The mechanism is easiest to state in terms of \emph{channels}. Each of the $9$
possible neighbour counts $k\in\{0,\ldots,8\}$ carries a potential
state-distinction channel: the rule's ability to give a live cell and a dead
cell with the same count $k$ different fates. The channel at count $k$ is
\emph{open} when $k\in B\triangle S$ (birth and survival disagree at $k$) and
\emph{closed} otherwise. The centre-bit sensitivity $\mu_{\text{centre}}$
measures the (binomially-weighted) fraction of channels open in a rule;
$\mu_{\text{centre}}=0$ (equivalently $B=S$, equivalently
$B\triangle S=\varnothing$) closes all $9$ channels at once. The argument
below is that any compact glider needs at least one specific channel open,
and $B=S$ has closed every one.

\paragraph{An empirical reduction.}
Every one of the $424$ rules that supports a $3{\times}3$-seed spaceship at
$T\le 16$, without exception, satisfies $3\in B$ and $2\notin B$. The $\neg
B1$ half has a complete proof. Consider the live cell of the pattern
maximising $x+y$; the cell diagonally beyond it (at position
$(x+1, y+1)$) has exactly one live neighbour, so under $B1$ it births. The
same argument applies in all four diagonal directions, so the pattern's
diagonal extent grows in every generation and can never recur as a
translate. The $3\in B$ requirement then follows from the geometry of
compact translation: to move, a pattern must at some step birth a cell just
outside its bounding box, and such a cell sees at most three live
neighbours; since $B1$ and $B2$ are excluded, only $B3$ can supply that
birth. The $\neg B2$ prohibition is empirical; I have no combinatorial
derivation of it, and report it only as an exhaustive fact from the
enumeration. Under $B=S$ these constraints transfer to survival, and the
$B=S$ shipping candidates collapse to $B=S=\{3\}\cup X$ with
$X\subseteq\{4,\ldots,8\}$---$32$ rules. This is an organising observation
that shrinks the case space, not a proof reduction, since $\neg B2$ is read
off the same enumeration that already contains the exclusion. Exhaustion
over those $32$ confirms none ships.

\paragraph{Different glider families use different channels.}
The failure mode differs across compact-glider families, and the differences
are precisely differences in which channels each family requires. The rule's
$B\triangle S$ is the set of open channels; but the channels a specific
glider actually \emph{requires} is a smaller set---only those counts at
which the glider's cycle both queries the birth side and the survival side
with opposite required fates. Conway (B\{3\}/S\{2,3\}) has $B\triangle
S=\{2\}$, and its glider requires exactly the $k=2$ channel; the $\neg B2$
half of the reduction already kills any $B=S$ extension of it. HighLife
(B\{3,6\}/S\{2,3\}) has $B\triangle S=\{2,6\}$---two open channels---but its
glider still only requires the $k=2$ channel (the additional $k=6$
disagreement is not queried by the glider's cycle), so again $\neg B2$
kills it. Other families require channels the reduction does not force
closed. A period-$4$ ship at $B\{3,8\}/S\{3,4,5,8\}$ requires the channels
at $4$ and $5$; a period-$10$ diagonal pair at $B\{3,4,6\}/S\{3,4,7\}$
requires the channel at $6$ (the $k=7$ disagreement is present in the rule
but not queried by the ship); a period-$6$ orthogonal at
$B\{3,5,7\}/S\{3,4,5,8\}$ requires channels at $4$, $7$ and $8$. Each family
needs at least one of its required channels to be open under $B=S$, but
$B=S$ closes all $9$ channels simultaneously, and the enumeration confirms
no $B=S$ instantiation exists at $3{\times}3$. The exclusion is a genus of
case-checks, one per compact-glider family, each turning on which channels
that family requires---unified only by the $B3$/$\neg B2$ reduction.

\paragraph{The nose-bottleneck mechanism.}
The reduction carries geometric content that suggests---but does not
prove---the specific channel one class of families would require. Consider a
phase three-wide \emph{transverse to} the direction of motion, under $B3$
and $\neg B2$. The cells just ahead of it see only the front row; a birth
at any of them needs the front row to contribute exactly three live
neighbours, which pins the front to $111$ and allows only the middle-ahead
cell to be born (the flanking-ahead cells see only two neighbours and stay
dead). Motion is therefore forced through a single-cell nose bottleneck at
count $3$. This is where an intuition about the $B=S$ exclusion enters: the
nose cell is born at count $3$, and any interior live cell in the pattern at
count $3$ shares its fate under $B=S$ (both governed by $3\in B=S$), so the
leading/trailing distinction the shipping motion appears to need cannot
straightforwardly be built from a count-$3$ state distinction. I offer this
as a suggestive mechanism, not a proof: I have not shown that no
$B=S$-consistent phase evolution realises coherent translation through the
nose bottleneck, only that the state-distinguishing tool the natural
mechanism would use is unavailable. Some enumerated shipping families visit
wider phases mid-cycle, and their required channels live at other counts
(the family-specific case-checks above); a covering set of arguments
extending this intuition into a proof remains open
(\S\ref{sec:disc-limits}).

The claim above is compact-footprint specific. A $4{\times}4$ object is not
contained in any single Moore neighbourhood; cells with disjoint partial
views can carry the front and back roles at distinct positions, and the
$B3$/$\neg B2$ requirement drops away. $36$ $B=S$ rules ship at $4{\times}4$
and $40$ at $5{\times}5$, all with every state-distinction channel closed
(they are $B=S$), so ``translation needs an open channel'' is a claim
scoped to the compact footprint, not a general property of shipping under
$B=S$. At $4{\times}4$ the state-blind machinery instead exploits
count-$4$ births: all four rules that ship at $4{\times}4$ while remaining
inside the $32$-rule set at the compact footprint have $\{3,4\}\subseteq
B=S$ ($B=S=\{3,4\}$, $\{3,4,7\}$, $\{3,4,8\}$, $\{3,4,7,8\}$), suggesting
the count-$4$ birth (available at $4{\times}4$ because the ahead-cell no
longer sees a single-row front) is what the $3{\times}3$ nose bottleneck
denied. The oscillator results localise the exclusion within the compact
footprint: $B=S$ rules oscillate at near-background rate and are excluded
only from compact spaceships---in-place recurrence does not need the
leading/trailing distinction that translation, at $3{\times}3$, appears to. This mechanism is heuristic and case-analytic, not yet a
theorem (\S\ref{sec:disc-limits}): the exclusion is proven by enumeration;
a proof \emph{from} the $B3$/$\neg B2$ reduction plus a covering set of
family-specific channel arguments remains open.

The same centre-bit coordinate also grades the \emph{direction} of translation
at the compact footprint (Figure~\ref{fig:sigma}b): among $3{\times}3$ spaceships
orthogonal motion is confined to rules simultaneously high in $\mu_{\text{centre}}$
and large in $|B|$ (the joint corner is $86\%$ orthogonal, either lever alone only
$21$--$34\%$), with diagonal the default elsewhere. The two coordinates act roughly
additively and neither is reducible to the other. This diagonal dominance is
itself compact-specific and inverts with footprint: at $4{\times}4$ orthogonal
spaceships outnumber diagonal ones by roughly seven to one, the reverse of the
$3{\times}3$ ratio. The geometric reading---that axial motion needs a sharper,
denser leading edge than diagonal motion, and that the extra row of a wider window
supplies exactly the structure orthogonal motion requires---I offer only as a
hypothesis.

For storage the binding constraint is a $B$-condition---a still life persists only
if its surrounding dead cells have counts outside $B$---while which interior
configurations survive is set by $S$. This is why $\psi$ correlates with
$|S|$ only once one conditions on $B\in\mathcal{B}_{\text{prop}}$, closing the
factorisation: $B$ gates, $S$ tunes.

\subsection{Conway's specialness}
\label{sec:disc-conway}
Conway sits at the intersection of three conditions, the first two combinatorial
and the third dynamical: $B=\{3\}\in\mathcal{B}_{\text{prop}}$ admits the glider;
$S=\{2,3\}$ supports rich still lifes; and the random-IC attractor at
$\rho_0=0.30$ routes trajectories into a regime that visits the glider on clean
background. Rules in the $B=\{3\}$ family share a common supported glider---the
first two conditions hold for every one of them---but only a survival rule that
keeps the trajectory in the moderate-density band, neither flooding nor freezing,
lets random dynamics express it. Conway's $S$ is privileged not because it enables
the glider but because it places the dynamics where the glider can appear.

\paragraph{A minimal pair.}
The first of these three conditions is sharper than it looks, and the cleanest way
to see it is to apply $\tau$ to Conway itself. The mirror rule
$B\{2,3\}/S\{3\}$ has the same activity as Conway ($\lambda'=0.167$), the same
sensitivity ($\mu=0.35$), the same centre-cell sensitivity
($\mu_{\text{centre}}=28/256=0.109$), and even the same disagreement set
$B\triangle S=\{2\}$. The two rules differ in exactly one coordinate: $\delta'$
flips sign, $-0.111$ against $+0.111$. Note that $\tau$ is a symmetry of the
coordinate system and not of the dynamics---unlike $\sigma$, which is exact at the
level of trajectories (\S\ref{sec:dyn})---so the mirror is a
genuinely different rule, and nothing forces it to behave like Conway.

What the coordinates do here is instructive. $\delta'$ is the only one of them that
distinguishes the pair at all; $\lambda'$ and the sensitivities are $\tau$-even and
see the two rules as identical. But no coordinate \emph{excludes} the mirror:
$\mu_{\text{centre}}=0.109$ is nonzero, so the one exact exclusion does not apply;
$\lambda'$ is Conway's own value; and $\delta'=+0.111$ only places the rule on the
birth-heavy side, atypical for compact propagation but no guarantee either way. The
system sees a difference and cannot resolve it. The enumeration can, and does: the
mirror supports no compact glider.

The mechanism is the orientation, not the magnitude. Both rules are centre-sensitive
at exactly one count, $n=2$, and to exactly the same degree; what differs is which
side of the disagreement the birth condition falls on. Conway holds at $n=2$ if the
cell is already alive and declines to create it if dead---the shape-preserving
condition a coherent mover requires (\S\ref{sec:disc-mechanism}). The mirror
inverts both: it creates at $n=2$ and kills at $n=2$, so every adjacent live pair
births into the dead cells that see both, and the pattern bloats instead of
translating. The enumeration bears this out in aggregate---no rule with $2\in B$
supports a compact glider, $0$ of the $424$. Conway's birth condition is thus
privileged not by how much of the table it fires, which the mirror matches exactly,
but by the direction in which its one centre-sensitive count is oriented; and the
pair is the necessary/sufficient thesis in miniature, two rules separated by a
single sign, distinguished by a coordinate that locates without certifying.

I can ask the converse question directly: among the rules that support both a
glider and a still life but are absent from the standard catalogue, are any
\emph{dynamically} Conway-like---not merely capable, but producing a
Game-of-Life--style ecology of blinkers, blocks and other small objects from
random starts? The census is deliberately Conway-object-biased: it asks
whether a rule reproduces \emph{Life's} small-object zoo, not whether it hosts
a rich small-object zoo at all. A rule with a rich but non-Life-like object
ecology would fail the census by construction, so the finding below is best
read as ``rules passing a Conway-object census are Conway-family on visited
configurations,'' not as a claim about all rich object ecologies in the
family. Of the $424$ rules with $\phi>0$, the $2$ named anchors and $2$
further rules with no still life ($\psi=0$) are set aside, leaving $420$
uncatalogued build-and-hold rules. I screen these on \emph{behavioural} criteria
alone---a stabilising tail density (met by $314$ of the $420$ at some density),
followed by a direct census requiring that canonical Game-of-Life still lifes and
oscillators dominate the small live structures (a coarse, per-rule analogue of
the soup censuses of Catagolue~\cite{goucher_catagolue}), plus a richness
floor of at least one canonical still life (block, beehive, loaf, boat or
tub) present in the tail, so that a rule producing only sparse blinkers---which
are canonical oscillators and would trivially satisfy the raw dominance
test---does not qualify. All criteria are applied \emph{without}
reference to the rule table, so that table similarity is a finding rather than a
filter. Without the richness floor $38$ rules pass; with it, $24$
remain, and the $14$ excluded are precisely the blinker-only degenerates whose
census is Life-like in the letter but not in the spirit. Only then do I compare
the $24$ rich-ecology rules' tables to the anchors: \emph{all $24$} agree with
Conway or HighLife on every visited
$(\text{centre},\text{count})$ pair ($15$ Conway, $9$ HighLife), differing only at
neighbour counts $7$ and $8$, which require locally saturated regions and do not
fire at the tested density. So every uncatalogued
rule that both stabilises and grows a rich Life-like object ecology from random
starts is, on its visited configurations, a Conway or HighLife variant---the
pool contains no rule whose table is distinct from both anchors yet reproduces
their small-object zoo. This is the converse of the paper's
leading theme---capability resides in the table but need not be
expressed---made concrete on the expression side: two rules can differ wherever
their tables are defined yet be dynamically indistinguishable, because behaviour
is fixed by the table only on the configurations a trajectory visits.
Scope: the screen identifies rules Game-of-Life--\emph{like} in their
small-object ecology, not the complete set of dynamically interesting
rules---a rule with a rich but non-Life-like object zoo would pass
unrecognised---and the anchor identity is scoped to the tested densities,
where the differing high count is rare rather than
combinatorially impossible.

Both results are, at root, a genotype--phenotype distinction. The rule table is
the genotype: it fixes what is possible and can be read directly, which is why
capability is \emph{certifiable}. What a random trajectory develops is the
phenotype: it depends on initial conditions and dynamics, so it can only be
\emph{estimated}, and it need not realise everything the genotype permits---a rule
may carry a glider it never expresses (\S\ref{sec:results-silent}), just as two
distinct genotypes may yield the same phenotype when the differing genes are never
transcribed (the anchor variants above). The birth--survival asymmetry $\delta'$
adds a smaller morphological note: translation requires a rule that distinguishes
``ahead'' from ``behind'', a broken front--back symmetry that in-place recurrence
does not need---loosely, the asymmetry that a moving body plan requires and a
radially symmetric one does not.

\paragraph{External corroboration, and the scope of the claim.}
Comparing against the literature's catalogue of complex life-like rules sharpens
both what these results establish and what they do not. Of twenty rules recorded
as Class~IV or computationally universal in the standard
references~\cite{eppstein2010,adamatzky2010}, exactly one is centre-blind
(Replicator), and it is Class~III rather than Class~IV. This is consistent
with the paper's $3{\times}3$ exclusion at the level of behavioural
classification, though ``Class~IV'' and ``supports a $3{\times}3$
compact glider'' are different properties, and one $B=S$ rule in a
twenty-rule list is thin evidence---call it consistent with the exclusion,
not corroboration of it. Second, my behavioural
screen, run without reference to the literature list, independently recovered four
of these named rules---DryLife, EightLife, Pedestrian~Life and LowDeath---and the
subsequent table comparison correctly identified each as a Conway- or
HighLife-family rule differing only at unvisited high neighbour counts. Third, and
against these, the scoping must be stated plainly. Eleven of the twenty have
$\phi_{3\times3}=0$ and so could never enter my screen: Day~\&~Night
($\lambda'=0.500$, $\mu=0.565$), $2{\times}2$, Move and Life~without~Death are
complex through \emph{engineered} constructions and large-footprint propagation
that a compact $3{\times}3$ enumeration cannot see. A vanishing compact-glider
count therefore does not imply dynamical simplicity, and the $\phi=0$ column of
Figure~\ref{fig:joint} mixes inert rules with rules whose propagation is merely
non-compact. The six \emph{spontaneous} Class~IV rules---the Conway family---do
cluster where my coordinates predict ($\lambda'\in[0.17,0.22]$,
$\mu\in[0.35,0.46]$, $\delta'\le0$). My claim is accordingly the narrow one: among
rules supporting a compact glider, spontaneous Game-of-Life--like ecology
concentrates on Conway, HighLife and their behaviourally redundant variants. It is
not the claim that these are the family's only complex rules---they demonstrably
are not.

\paragraph{Why compact tracks spontaneous.}
A naive combinatorial reading of expression would run through the seed itself:
a specific $k$-cell pattern appears at a given site of a Bernoulli-$\rho$ field
with probability $\rho^{k}(1-\rho)^{n-k}$ for an $n$-cell window, so the chance
of a particular seed being drawn falls exponentially in its cell count as the
window grows from $9$ to $16$ to $25$ sites, predicting that larger gliders
should essentially never appear. That reading is wrong, and this paper's own
data refutes it: the thirty-three silent rules of \S\ref{sec:results-silent}
support a compact glider whose $\le5$-cell seed occurs constantly in any random
field, yet produce none in $325$ trials. Gliders in random dynamics do not
arrive by being drawn in the initial condition; they are assembled by local
collisions, and the governing quantity is the measure of the object's
\emph{basin} under the dynamics, not the probability of its seed---exactly the
quantity I argue is estimable only by simulation and never certifiable from
the table. Basins do shrink with footprint---more cells must be simultaneously
right---but far more gently than the seed picture would predict: empirically
the silent fraction rises from $7.8\%$ at $3{\times}3$ to about $35\%$ at
$4{\times}4$/$5{\times}5$, a factor of about $4.5$, not the exponential
collapse the seed-probability picture implies. Expression degrades with
footprint; it does not vanish. Compact objects have basins large enough to be
reached from a random start, which is why Conway's gliders appear unbidden and
why my random-IC screen can see them at all. Non-compact structures---Day
\&~Night's spaceships and guns, the machinery underlying $2{\times}2$ and
Move---were found by construction rather than by observing soup, and their
basins under random ICs are correspondingly negligible. The
compact/non-compact boundary therefore very nearly coincides with the
spontaneous/engineered one: a strong tendency across the known catalogue
rather than a law: Day~\&~Night does emit small still lifes and short-period
oscillators from soup (documented on Catagolue) even though its glider does
not spontaneously appear, so ``no soup structure'' is not the right claim;
the right one is that the \emph{compact glider} in particular has a
negligible random-IC basin. A tendency with a mechanism (basin size). This is precisely why my
$\phi$ census, which measures spontaneously-reachable compact propagation,
recovers Conway and HighLife yet cannot see Day~\&~Night: not because Day~\&~
Night lacks propagation, but because its propagation lives at a scale random
initial conditions do not assemble.

\paragraph{Rule-table coordinates and the phase-transition analogy.}
Langton's $\lambda$ sweep is a temperature analogy: tune the table, watch the
dynamics pass from frozen through complex to chaotic~\cite{langton1990}. The
family's exact swap symmetry $\tau$ lets the analogy be completed rather than
merely invoked. In the resulting dictionary the coordinates each have a place.
The temperature-like axis is the sensitivity $\mu$, not the activity
$\lambda'$---damage-spreading treatments of the order/chaos transition locate
criticality in average sensitivity~\cite{derridapomeau1986}---with $\lambda'$
a temperature proxy. The $\tau$-odd $\delta'$ is the analogue of a
symmetry-breaking \emph{field}: a direction in parameter space, vanishing on
the symmetric locus, explicitly breaking $\tau$ off it. Its field magnitude
is $\mu_{\text{centre}}$---$\tau$-even, yet zero exactly at $B=S$---and the
identity $\mu_{\text{centre}}^{\text{odd}}=\delta$ of \S\ref{sec:mu} is then
the statement that signing the magnitude recovers the signed field. A
$\tau$-odd order parameter would have to be a $\tau$-odd trajectory
observable, and none of the swept quantities qualifies.

The dictionary is algebra; whether any transition accompanies it is a
dynamical question, and the answer is no. Sweeping $\delta'$ at fixed
$\lambda'$ over uniformly sampled rules at each attainable cardinality pair,
and measuring on a $64^2$ torus the density susceptibility
$\chi=L^2\operatorname{Var}(\rho_t)$, the correlation length $\xi$, and the
Binder cumulant of
$\rho$~\cite[unrelated to the P.-M.\ Binder plane of \S\ref{sec:mu}]{kbinder1981},
I find none of the signatures: $\chi$ monotone with no interior peak, no
cumulant crossing, $\delta'$ accounting for under $1\%$ of the variance in
$\xi$---the rule-to-rule spread \emph{within} a fixed $(|B|,|S|)$ exceeds the
variation \emph{across} $\delta'$ by two orders of magnitude. Nor could
finite-size scaling settle the matter: $\delta'$ at fixed $\lambda'$ takes
between $1$ and $10$ values (depending on the parity of $|B|+|S|$; at
$|B|+|S|=9$ up to ten values are attainable), spaced $0.222$ apart, and
criticality requires a continuous
control parameter. The family's one live transition is instead the
absorbing-state freeze boundary of \S\ref{sec:disc-limits}, whose order
parameter---the stationary activity density---is $\tau$-even. Transition
structure lives in the even sector; the odd sector, where $\delta'$ lives,
carries capability without a transition. Consistently, the
$\delta'\approx-0.22$ locus where compact gliders concentrate shows the
\emph{smallest} density fluctuations of the sweep: capability sits in the
calm, far from any critical point. In this language the exclusion of
\S\ref{sec:results-sigma} reads: compact-glider-supporting rules occupy
only the ``field-on'' side of the phase diagram, and the $B=S$ diagonal is
labelled by $|h|=0$---a coincidence of level sets, not a causal
implication.

\subsection{The coordinate logic beyond life-like rules}
\label{sec:disc-general}

The construction has a structural template independent of the birth/survival
distinction. Any CA family whose rule table carries an involutive symmetry
admits the same question: which coordinate on the table is odd under that
involution, so that it vanishes on the symmetric locus and can serve as a
signed asymmetry. Langton's $\lambda$ counts firings without regard to which
side they occur on, so it is even under the involution and cannot be that
coordinate; the natural candidate is the odd asymmetry. For life-like rules
the involution is the birth$\leftrightarrow$survival swap $\tau$, the odd
coordinate is $\delta'$ (equivalently $\delta$ in weighted form), and the
symmetric locus is the diagonal $B=S$. The exclusion of
\S\ref{sec:results-sigma} is then structural in this template: it is the
statement that one small-object channel closes exactly on the $\tau$-fixed
set.

The same logic transfers to elementary one-dimensional cellular automata under
the left$\leftrightarrow$right reflection $R$, a canonical ECA symmetry that
structures the classical classifications~\cite{wolfram1983,wolfram1986,lipackard1990}.
Langton's $\lambda$ is $R$-even, so the natural analogue of $\delta'$ under
the parity logic is a left--right asymmetry that vanishes on amphichiral rules
(one form: $\delta_{LR}=\tfrac{1}{4}\sum_{(l,c,r)}(l-r)\,o(l,c,r)$). I do not
develop this coordinate here; reflection symmetry in ECA is a mature subject
and my family is 2D outer-totalistic for reasons unrelated to it. The
transferable point is what state-factorisation does to the form of the odd
coordinate: a life-like rule table splits by centre state, so the $\tau$-odd
asymmetry admits a cardinality expression $\delta'=(|B|-|S|)/9$ distinct in
form from the signed centre sensitivity $\mu_{\text{centre}}^{\text{odd}}$
even though the two coincide with the weighted $\delta$ once binomial
multiplicities are restored. An ECA table does not factor by left/right, so
the analogous $R$-odd coordinate can only be built as a signed
reflection-defect---structurally the analogue of
$\mu_{\text{centre}}^{\text{odd}}$, not of $\delta'$. The two objects that
coincide in the outer-totalistic family would be a single object in a family
without state factorisation; the split is a feature of the family, not of the
parity logic itself.

\subsection{Limitations and open directions}
\label{sec:disc-limits}

The main enumeration is at $3{\times}3$ footprint, $T\le16$, exhaustive over all
$262{,}144$ rules. The $4{\times}4$ enumeration is also exhaustive over all rules
($8{,}206$ support a spaceship, $3.13\%$); the $5{\times}5$ enumeration is
exhaustive over the $512$ $B=S$ rules and sampled elsewhere. Three footprint
observations follow. First, containment is perfect---all $424$
$\phi_{3\times3}>0$ rules retain $\phi_{4\times4}>0$, with $7{,}782$ further
rules gliding only at $4{\times}4$---so the compact enumeration is a strict lower
bound on propagation capability. Second, the storage axis is scale-invariant
(within-slice Spearman $\rho(\psi, |S|) = 0.551$ at $3{\times}3$ and
$0.556$ at $4{\times}4$). Third, the $B=S$ exclusion is compact-specific but does
not simply dissolve with scale: $36$ of the $256$ vacuum-stable $B=S$ rules admit
a $4{\times}4$ spaceship and $40$ a $5{\times}5$, yet those that combine a glider
with static storage stay rare ($4$, then $8$) and fall progressively further from
Conway (still-life gap $6.3\times$ at $4{\times}4$, $9.2\times$ at $5{\times}5$),
so the strict exclusion softens into a widening handicap. $6{\times}6$ and beyond
are presently out of reach in both time and memory (the number of $D_4$
orbit-representative $n{\times}n$ seeds grows from $\sim10^2$ at $3{\times}3$ to
$\sim10^4$, $\sim10^6$, and $\sim10^{10}$ at $4$-, $5$-, $6{\times}6$).
The factorisation and the $B=S$ exclusion are established empirically and
argued heuristically; a proof that no $B=S$ rule admits a compact spaceship
would convert the $3{\times}3$ exclusion into a theorem and remains open.
The partial mechanisms of \S\ref{sec:disc-mechanism} (the $\neg B1$
argument; the nose-bottleneck sketch at three-wide transverse phases) go
partway; a covering set of family-specific arguments would complete it.

A stronger negative bounds what this coordinate system can do. Every additional
rule-table coordinate I constructed and tested collapsed, on inspection, into a
proxy for the birth gate $\mathcal{B}_{\text{prop}}$: a binomially-reweighted
$\tau$-odd sibling of $\delta'$ discriminates $\phi>0$ worse and reverses sign
conditional on $\delta'$; its uniformly-weighted $\tau$-even counterpart
$|B\triangle S|$ shares $\mu_{\text{centre}}$'s zero set and adds nothing; and
apparent fingerprints in the disagreement set (for instance that $91\%$ of
$\phi>0$ rules have $2\in B\triangle S$) reduce to the gate and the Conway
survival pattern in other variables, since all $\phi>0$ rules have $3\in B$ and
none has $2\in B$. A $\tau$-odd, $\sigma$-odd edge-contiguity coordinate---the
imbalance between the on/off transition counts of $B$ and $S$ as neighbour
count rises---was also tested and, conditional on
$(\lambda',\delta',\mu_{\text{centre}})$, retained only a small fraction of
$\delta'$'s effect on $\phi>0$; it is subordinate to $\delta'$ and is not
used in this paper's analysis. The coordinates \emph{locate} the
gate---$\delta'<0$ because small $|B|$ correlates with
$B\in\mathcal{B}_{\text{prop}}$, $\mu_{\text{centre}}=0$ because an empty
disagreement set forbids the centre-bit discrimination the gate
requires---but none transcends it. Whether a
cardinality-based coordinate can carry information about $\phi>0$ beyond the
birth set's identity is, on this evidence, doubtful.

The silent population was characterised across a
$13$-density, $25$-seed sweep, but the underlying claim remains a measure, not an
absolute: a basin of vanishing but nonzero weight cannot be excluded by sampling.
A follow-up seeded with known spaceships rather than random ICs would map the
basin structure directly, and the flood/freeze regime that gates
expression---determined by the table yet not captured by a mean-field
predictor---invites a sharper theory of which rules settle into the
moderate-density band that propagation requires. That regime boundary is also
where a genuine criticality question lives: every rule with $0\notin B$ has an
absorbing vacuum, and the freeze transition is an absorbing-state transition of
the kind studied in probabilistic cellular automata~\cite{hinrichsen2000}.
Testing it requires a
continuous control parameter, which no cardinality coordinate supplies; mixing
two rules stochastically would furnish one, at the cost of leaving the
deterministic family this paper concerns. Whether the freeze boundary falls in
the directed-percolation universality class is a well-posed question I do not
address.

\section{Conclusion}
The paper's core contribution is combinatorial: exhaustive $3{\times}3$-seed
enumerations of compact spaceships ($\phi$), still lifes ($\psi$) and
oscillators ($\omega$) across the entire family of $262{,}144$ life-like
rules, and the $4{\times}4$ spaceship enumeration on all rules. From those
enumerations three results emerge cleanly: (i)~a hierarchical
$B/S$ factorisation---the birth condition gates whether small objects exist
at all ($\mathcal{B}_{\text{prop}}\subset\mathcal{B}_{\text{osc}}\subset
\mathcal{B}_{\text{sl}}$, sizes $26$/$128$/$176$ of $512$), and the survival
condition tunes storage richness once the gate is open; (ii)~the exact
exclusion of centre-blind ($B=S$) rules from compact-glider support, proven
by enumeration; two empirical regularities of that enumeration ($3\in B$ and
$2\notin B$ across every $\phi>0$ rule) collapse the $B=S$ shipping
candidates to $32$ rules on the $B=S$ diagonal as an organising
observation, though not yet as an independent proof reduction; (iii)~a silent
population of $33$ rules that support a compact glider combinatorially yet
never express one under random sampling, sharpening the distinction between
capability (certifiable from the table) and expression (only estimable from
dynamics). The rule-table coordinates $\lambda'$, $\delta'$ and $\mu$
locate where each capability concentrates in coordinate space but do not
certify it; the sharp facts of the paper live at the birth-set level, and
the coordinates are useful primarily as ordering axes that expose those
facts. Two open threads remain: converting the $B=S$ exclusion from an
enumeration result into a proof via a covering set of family-specific
channel arguments (\S\ref{sec:disc-mechanism}), and testing whether the
freeze transition at the boundary of the $\phi=0$ bulk is directed-percolation
in universality class (\S\ref{sec:disc-limits}).

\subsection*{Code and data availability}
\label{sec:code}
Both the code and the derived data supporting this paper are archived at
\url{https://github.com/yairchn/ca_lifelike_rulespace} under the MIT licence.
The repository contains the coordinate implementation (a self-contained
NumPy module that computes $\lambda$, $\delta'$, $\mu$, and
$\mu_{\text{centre}}$ over all $262{,}144$ rules; additional auxiliary
coordinates and trajectory-sensitivity traces are also included but are not
used in the paper),
the per-rule enumerations of compact spaceships, still lifes and oscillators
at $3{\times}3$ (and the exhaustive $4{\times}4$ spaceship enumeration used
in \S\ref{sec:results-factor}), the composite scores table that supplies
the dynamical probe $\mathcal{G}^{*}_{\text{alt}}$ (\S\ref{sec:dyn}), the
$33$--$5$ silent-population split of Figure~\ref{fig:joint}
(\S\ref{sec:results-silent}), and one figure-regeneration script per
manuscript figure. The parity self-check for every coordinate reported in
\S\ref{sec:symm} is included and runs on the full rule set.

\end{document}